%% file: main.tex
\documentclass[12pt, a4paper, twoside,authoryear,sort]{scrartcl}
\input{preamble.tex}
\usepackage{multirow}
\newcommand{\beginsupplement}{%
        \setcounter{table}{0}
        \renewcommand{\thetable}{S\arabic{table}}%
        \setcounter{figure}{0}
        \renewcommand{\thefigure}{S\arabic{figure}}%
     }
     
\newcommand{\beginappendix}{%
        \setcounter{table}{0}
        \renewcommand{\thetable}{A\arabic{table}}%
        \setcounter{figure}{0}
        \renewcommand{\thefigure}{A\arabic{figure}}%
     }     

\begin{document}
	\input{0-title}
	\input{1-introduction}

    \input{3-DAG}
     \input{2-Dvineregression}
     \input{4-vine-analysis}

\input{5-validation}
     \input{6-future-directions}

    \input{7-Appendix}

\end{document}

%% file: preamble.tex
\pdfoutput=1

\usepackage[tracking=true,expansion=true,stretch=15,shrink=15]{microtype}
\DeclareMicrotypeSet*[tracking]{my}
{ font = */*/*/sc/* } 
\SetTracking{encoding = *, shape = sc }{ 45 }
\SetProtrusion{encoding={*},family={bch},series={*},size={6,7}}
{1={ ,750},2={ ,500},3={ ,500},4={ ,500},5={ ,500},
	6={ ,500},7={ ,600},8={ ,500},9={ ,500},0={ ,500}}
\SetExtraKerning[unit=space]
{encoding={*}, family={qhv}, series={b}, size={large,Large}}
{1={-200,-200}, 
	\textendash={400,400}}

\usepackage{setspace}
\usepackage[
	left=1.1in,
	right=1.1in,
	top=0.7in,
	bottom=0.7in,
	includeheadfoot
]{geometry}
\usepackage[automark, headsepline]{scrlayer-scrpage}

\usepackage[disable]{todonotes}
\usepackage{lineno}
\usepackage{amssymb}
\usepackage{amsmath}
\usepackage{amsthm}
\usepackage{dsfont}
\usepackage{bm}
\usepackage{subfig}
\usepackage{enumerate}
\usepackage{graphicx}
\usepackage{float} 
\usepackage{tikz}
\usetikzlibrary{shapes,arrows,decorations.pathmorphing,graphs,positioning,backgrounds}
\usepackage[round,sort,comma]{natbib}
\usepackage{hyperref}
\usepackage{nameref}
\definecolor{TUMblue}{cmyk}{1, .54, .04, .19}
\hypersetup{
	colorlinks   = true, 
	urlcolor     = TUMblue, 
	linkcolor    = TUMblue, 
	citecolor   = TUMblue 
}

\newtheoremstyle{new}{12pt}{12pt}{\itshape}{}{\bfseries}{.}{1em}{}
\theoremstyle{new}

\newcommand*\linenomathpatchAMS[1]{%
	\expandafter\pretocmd\csname #1\endcsname {\linenomathAMS}{}{}%
	\expandafter\pretocmd\csname #1*\endcsname{\linenomathAMS}{}{}%
	\expandafter\apptocmd\csname end#1\endcsname {\endlinenomath}{}{}%
	\expandafter\apptocmd\csname end#1*\endcsname{\endlinenomath}{}{}%
}

\expandafter\ifx\linenomath\linenomathWithnumbers
\let\linenomathAMS\linenomathWithnumbers
\patchcmd\linenomathAMS{\advance\postdisplaypenalty\linenopenalty}{}{}{}
\else
\let\linenomathAMS\linenomathNonumbers
\fi

\newcommand{\fX}{{\bm X}}

\newcommand{\fx}{{\bm x}}

\newcommand{\fu}{{\bm u}}

\allowdisplaybreaks

%% file: 0-title.tex
\title{Analysis of an interventional protein experiment using a vine copula based structural equation model}

\author{Claudia Czado\footnote{Corresponding author, Department of Mathematics, Technical University of Munich, Boltzmannstr. 3, 85748 Garching, Germany
(email: \href{mailto:cczado@ma.tum.de}{cczado@ma.tum.de})} ~and Sebastian Scharl \footnote{Department of Mathematics, Technical University of Munich, Boltzmannstr. 3, 85748 Garching, Germany
(email: \href{mailto:sebastian.scharl@gmail.com}{sebastian.scharl@gmail.com})}}

\date{\hspace{3pt} \normalsize\today}

\maketitle


\begin{abstract} 
\noindent {\bfseries \sffamily Abstract}\\
While there is considerable effort to identify signaling pathways using linear Gaussian Bayesian networks from data, there is less emphasis of understanding and quantifying conditional densities and probabilities of nodes given its parents from the identified Bayesian network. Most graphical models for continuous data assume a multivariate Gaussian distribution, which might be too restrictive. We reanalyse data from an experimental setting considered in \cite{Sachs2005} to illustrate the effects of such restrictions. For this we propose a novel non Gaussian nonlinear structural equation model based on vine copulas. In particular the D-vine regression approach of \cite{kraus2017d} is adapted. We show that this model class is more suited to fit the data than the standard linear structural equation model based on the biological consent graph given in \cite{Sachs2005}. The modelling approach also allows to study which pathway edges are supported by the data and which can be removed. For data experiment $cd3cd28+aktinhib$ this approach identified three edges, which are no longer supported by the data. For each of these edges a plausible explanation based on underlying the experimental conditions could be found. 
\end{abstract}


\pagestyle{scrheadings}
\clearscrheadings
\lohead{}
\rohead{\pagemark}
\lehead{Czado and Scharl}
\rehead{\pagemark}

%% file: 1-introduction.tex
\section{Introduction}

One important task in biology is to identify signalling pathways caused using a stimulus defined by an interventional experimental setting. The resulting cascade of chemical reactions on proteins in a cell has to be understood and quantified. A well known and publicly available data set for this kind of set up is the data collected and initially analyzed by \cite{Sachs2005}. It involves measurements on $d=11$ proteins on $n$ cells under different experimental conditions. 
Often the underlying pathways are learned from these cell measurements using graphical models on directed acyclic graphs
(DAGs), which are also called Bayesian networks \citep{Pearl1988,Lauritzen1996,KollerFriedman2009}. Here the associations between the $d$ random variables $X_1,\ldots, X_d$ are modelled by a  statistical model on a graph $\cal G$ with node set $V=\{X_1,\ldots X_d\}$ and edge set $E \subset V \times V$. Using Markov assumptions and allowing only for acyclic graphs, the joint density of $X_1,\ldots, X_d$  can be expressed as product of conditional densities
\begin{equation}
\label{eq:decomp}
f(x_1,\ldots x_d)= \prod_{i=1}^d f(x_i|\bm \pi(X_i)=\bm \pi(x_i)),
\end{equation}
where $\bm \pi(X_i)$ is the parent set of $X_i$, i.e the set of nodes which have an edge going to $X_i$. More precisely $\bm \pi(X_i)=\{X_j: X_j \rightarrow X_i\}$. Note that $\bm \pi(X_i)$ can be the empty set. The set at observed values of the parent nodes is denoted by $\bm \pi(x_i)$.

 \cite{Sachs2005} used a single DAG to model the pooled data collected under all experimental conditions.  Since the Sachs data involves only continuous data, Gaussian DAGs or  undirected Gaussian graphical models \citep{dempster1972covariance} are often utilized. Gaussian DAG models are often constructed using a structural equation model (SEM)
\citep{mulaik2009linear,kaplan2008structural}.

Many approaches have been followed to learn the underlying pathway network from this data. To name a few,
\cite{friedman2008sparse} used graphical lasso to construct an undirected graph for the pooled data, while \cite{ellis2008learning} discretized the data and used a graphical model for discrete data.  \cite{peterson2015bayesian} constructed a selection prior for different experimental conditions to conduct a joint Bayesian MCMC analysis. 
Earlier \cite{luo2011bayesian} developed a non graphical Bayesian hierarchical approach, where the regression coefficients for each experimental condition are linked by a prior. \cite{scutari2013identifying} used bootstrapping methods to identify significant edges for the Sachs data.  A penalized  maximum likelihood estimator of the interventional Markov equivalence classes under linear Gaussian assumptions was derived in \cite{hauser2015jointly}.  \cite{castelletti2018learning,castelletti2019objective} use an objective Bayes approach to learn the Markov equivalence classes in a Gaussian DAG for the Sachs data. \cite{ramsey2018fask} introduced a fast adjacency
skewness (FASK) algorithm, which orients edges derived from a fitted undirected network by incorporating skewness properties of the variables.  Genetic algorithms have also been applied to this data set (\cite{jose2020bayesian}).

Many of the pathway learning approaches assume that there is an underlying joint Gaussian model available which allows the conditional densities in \eqref{eq:decomp} as Gaussian. However this might not be true for the data considered. 
Already
\cite{hauser2015jointly} mentioned that the Gaussian assumption for the Sachs data is questionable. This was also noticed by \cite{voorman2014graph,zhang2017mixture}.  \cite{voorman2014graph} developed a nonparametric approach to include non Gaussian behavior in a graphical model, while \cite{zhang2017mixture}
 build a Bayesian network using a mixture copula on $d$ dimensions to model the conditional distribution of a node $X_i$ given the observed values of the  parents $\bm\pi(X_i)=\bm\pi(x_i)$ to accommodate the non Gaussianity of the data and the pooling over the different experimental conditions. Here he followed the approach of \cite{elidan2010copula}, who was the first to use a copula based representation of the conditional density $f(x_i|\bm \pi(X_i)=\bm \pi(x_i))$ in \eqref{eq:decomp}.
 
\begin{figure}
\centering
    \centering
    \subfloat[All edges]{{\includegraphics[width=0.45\textwidth]{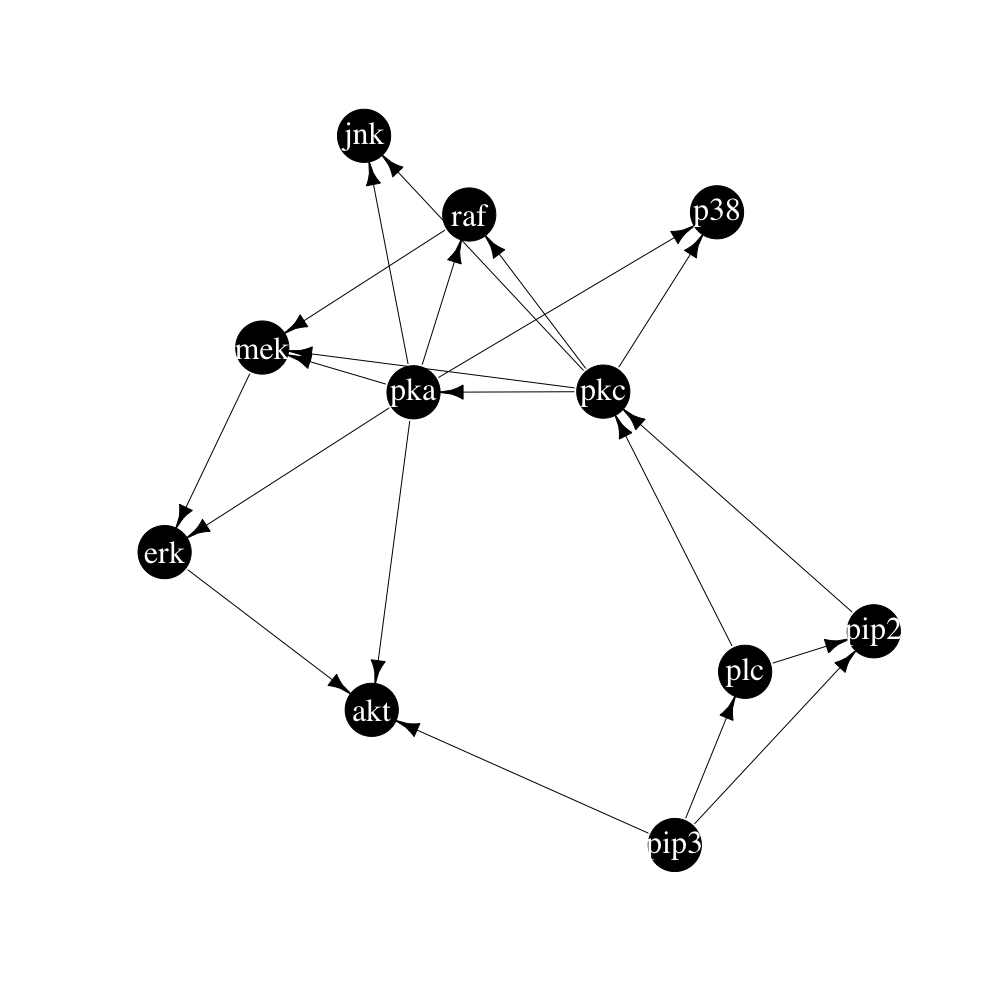}}}%
    \qquad
    \subfloat[Edges in the D-vine models]{{\includegraphics[width=0.45\textwidth]{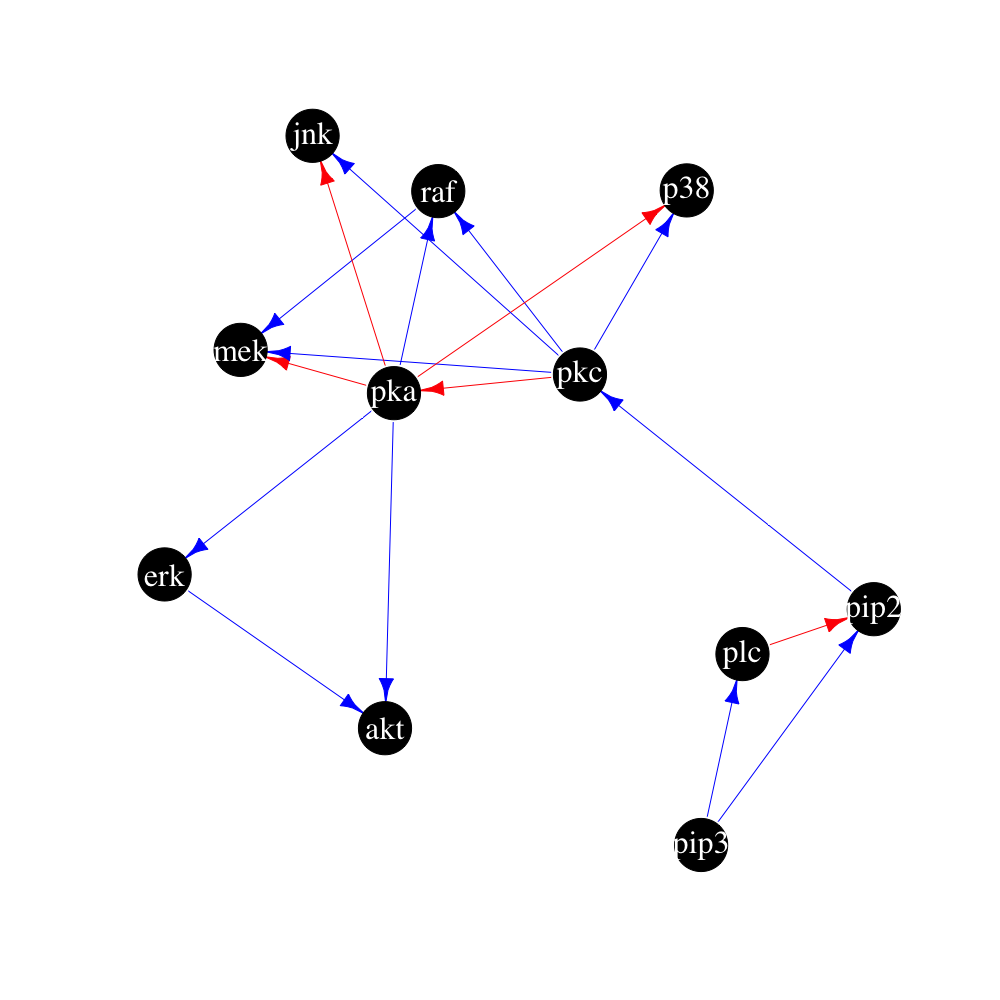}}}%
\caption{Panel (a): Consent DAG of the Sachs data. Panel (b): Modelled edges in all non Gaussian D-vine based SEMs in blue and red. Edges  missing in the Gaussian SEM are colored in red}
\label{fig:DAG}
\end{figure}

 One special feature of the Sachs data is that there exist a biological consent graph given in Figure 3 of \cite{Sachs2005}. It consists of 20 edges connecting the eleven nodes of the Sachs data. This consent graph is shown in  Panel (a) of Figure \ref{fig:DAG}. 
 
 In this paper we want to extend the flexibility by modeling the conditional densities $f(x_i|\bm\pi(X_i)=\bm \pi(x_i))$ in Equation \eqref{eq:decomp} using a copula based approach. For this we suggest the use of vine copulas (\cite{joe2014dependence,czadobook}), which have become very popular among practitioners. One major reason for this is the enormous flexibility of this class compared to standard multivariate copula families such as the class of elliptical or Archemedian copulas. In particular the vine copula class is constructed using only bivariate copulas, which can be chosen separately and data driven. For the conditional densities needed in the DAG model we use the ones derived from a fitted vine distribution of the node $X_i$ and its parents $\bm\pi(X_i)$. The vine distribution is chosen in such a way that the conditional density of a node given its parents is available in closed form. The choice of the vine model is further optimized such that the conditional log-likelihood is maximized when parent nodes are included in a forward selection.  For this we use the approach taken in \cite{kraus2017d} derived for vine copula based quantile regression models. The forward selection of the parent nodes will also be used to identify which edges in the consent graph are supported by the data and allows for a potential reduction of the edges in the graph.

 Overall this  approach allows us to fit very general vine copula based Bayesian networks, thus extending the copula Bayesian network of \cite{elidan2010copula}. We call the resulting models vine copula based SEMs and show the
  need of this novel model class for one experimental setting of the Sachs data using the consent graph. We restrict to a single experimental setting to avoid accommodating clusters present in the pooled data set arising from different experimental conditions.
 
 This new approach allows us to quantify  the effects of ignoring the non Gaussianity in the data with regard to the changes in the estimated conditional densities given the parent nodes as well as on simulated conditional quantiles based on the fitted models.
 In summary this extension provides an important building block for learning non Gaussian networks. It is easy to fit to a given graph, can remove nonsignificant edges in the graph and allows to quantify differences in conditional densities as well as probabilities. For one experimental setting of the Sachs we show that this approach leads to a better understanding and quantification of the graphical structure.
 
 The paper is organized as follows: In Section 2 we introduce statistical models on DAGs  including Gaussian DAGs more formally. The needed background on copulas 
 and especially on vine copulas is given in Section 3. This includes the subclass of D-vine copulas. General vine copula based regression models are discussed in Section 4. The included D-vine regression forms the building block of the novel D-vine copula based SEMs introduced in Section 5. In Section 6 we present D-vine copula based structural equation analysis for the  experimental setting $cd3cd28+aktinhib$ of the Sachs data. In Section 7 we provide simulation based evidence for the better fit of the D-vine copula based SEM compared to the standard linear Gaussian Bayesian network fit. In Section 8 we quantify differences between the fitted models based on conditional probabilities and densities. The paper closes with a summary and an outlook to future research directions.

%% file: 3-DAG.tex
\section{Statistical models on DAGs}
To describe the behavior of many variables it is important to identify how the variables interact. For many applications it is useful to describe these interaction by a network with directed edges.
Following \cite{Pearl1988} a Bayesian network is a directed acyclic graph (DAG), where each node represents a random variable $X_1,...,X_d$. The edges connecting the nodes indicate direct causal influence between the linked variables and their strengths are quantified by conditional probabilities or densities in the case of continuous random variables.  

Since we are dealing with continuous data in this paper we start with a linear SEM. For this we assume that we have a  graph $\cal G$ for the node set $V=\{X_1,\ldots X_d\}$ and edge set $E \subset V \times V$ with directed weighted edges. Let $A$ be an associated $d\times d$ dimensional adjacency matrix with $(i,j)$th element given by weight $A_{i,j}$, which will be estimated. Further $A_{i,j} \neq 0$ if and only if $(i,j)\in E$, i.e. there is an edge $i \rightarrow j$. Let $\bm\epsilon$ be a $d$ dimensional multivariate normal random vector with zero mean vector and diagonal covariance matrix $\Omega$. Then a linear SEM for $\bm X=(X_1,\ldots,X_d)^\top$ is defined as
\begin{equation}
    \label{eq:sem}
    \bm X= A^\top \bm X + \bm \epsilon. 
\end{equation}
It follows that $\bm X $ is multivariate normal with zero mean vector and covariance matrix $\Sigma$ satisfying $\Sigma^{-1}= (I-A) \Omega^{-1} (I-A)^T$. Using conditioning arguments the factorization  \eqref{eq:decomp} follows, where the the conditional densities are univariate normal densities. In general such a factorization is equivalent to the Markov assumption with respect to the graph $\cal G$ (see Theorem 3.27 of \cite{Lauritzen1996}). 

Since the graph is acyclic there is at least one permutation $\bm\eta=(\eta(1),\dots \eta(d))^\top$ of the numbers 1 to $d$ such that $A_{i,j}=0$ for all $\eta(i) \geq \eta(j)$. Such a permutation is called a topological order. One such topological order resulting from the consent graph in Figure \ref{fig:DAG}
applied to the eleven protein variables on the logarithmic scale $$\bm x=(pip3,plc , pip2, pkc, pka, p38,  jnk,  raf, mek, erk ,akt)^\top$$ of the Sachs data
is given in Table \ref{tab:TopOrd}.




\begin{table}
\centering
\begin{tabular}{|r|r|r|r|r|r|r|r|r|r|r|r|}
  \hline
  $\eta(1)$ & $\eta(2)$ & $\eta(3)$ & $\eta(4)$ & $\eta(5)$ & $\eta(6)$ & $\eta(7)$& $\eta(8)$ & $\eta(9)$ & $\eta(10)$ & $\eta(11)$ \\ 
  \hline \hline
  pip3 & plc & pip2 & pkc &  pka & p38 &   jnk &  raf & mek & erk & akt \\ 
   \hline
\end{tabular}
\caption{Topological order of the consent graph of  Figure \ref{fig:DAG}}
\label{tab:TopOrd}
\end{table}
Reordering the columns and rows of $A$ according to the topological order $\bm \eta$ allows to express A as a strict upper triangular matrix and thus \eqref{eq:sem} can be expressed recursively using the topological order. 
Utilizing \eqref{eq:decomp} and the topological order we can therefore express the joint density of the Sachs data with the eleven protein variables $\bm x$ based on the consent graph as
\begin{align}
\label{eq:LGBNDicht}
f(\bm x) 
=& 
f(pip3) * f(plc|pip3) * f(pip2|plc,pip3) *
f(pkc|pip2,plc) \nonumber \\
*&
f(pka|pkc) * f(p38|pka,pkc) * f(jnk|pka,pkc) * f(raf|pka,pkc)\\
*&f(mek|raf,pka,pkc) *f(erk|mek,pka) *
f(akt|erk,pka,pip3). \nonumber
\end{align}

Hence instead of estimating one eleven dimensional density we can estimate eleven univariate conditional densities, where each of them has at most three conditioning variables. 
We use the special case of a linear Gaussian Bayesian network (LGBN) as a reference model, where the weights of the conditional densities derived from  the SEM \eqref{eq:sem} are  estimated. In particular it satisfies


\begin{equation}
\label{eq:LGBN}
    X_i|\bm\pi(X_i)=\bm\pi(x_i)\sim 
    N(\beta_{i0}+\bm{\beta_i}^\top \bm\pi(\bm x_i);\sigma_i^2)
\end{equation}
for regression parameters $\beta_{i0}, \bm{\beta_i}$ and residual variances $\sigma_i^2$ for $i=1,\ldots,d$. The corresponding maximized log-likelihoods as well as AIC and BIC statistics are given in Table \ref{tab:GaussModX3} for the experimental setting $cd3cd28+aktinhib$ of the Sachs data. We see that the protein $pip2$ contributes more to the overall fit compared with the other proteins, while $akt$ does the opposite. This is expected in this experimental setting since we have activation of the levels for $pip2$ and inhibition for $akt$.

Since the SEM \eqref{eq:sem} and the LGBN \eqref{eq:LGBN} induce marginal normality an inspection of the marginal histograms 
of the Sachs experiment $cd3cd28+aktinhib$ shows skewed and multimodal histograms (see first column of Figure  \ref{fig:contALL3X3} of Appendix A1) thus indicating that the reference model might be inadequate. To extend beyond the LGBN we discuss now a copula based approach. This will allow us to construct nonlinear non Gaussian SEMs.

\begin{table}[ht]
\centering
\tiny
\begin{tabular}{|r|r|r|r|}
  \hline
 Node & log-likelihood &  AIC$_F$ & BIC$_F$  \\ 
  \hline    \hline
 pip3 & -986.90 & 1977.81 & 1987.29 \\ \hline
 plc & -921.77 & 1849.53 & 1863.75 \\ \hline
   pip2 & -1249.37 & 2506.74 & 2525.70 \\ \hline
   pkc & -952.32 & 1912.64 & 1931.60 \\ \hline
  pka & -830.92 & 1667.84 & 1686.80 \\ \hline
  p38 & -570.72 & 1149.45 & 1168.40 \\ \hline
   jnk & -991.87 & 1991.74 & 2010.70 \\ \hline
  raf & -969.82 & 1947.64 & 1966.60 \\ \hline
   mek & -551.50 & 1112.99 & 1136.69 \\ \hline
   erk & -968.07 & 1944.14 & 1963.09 \\ \hline
    akt & -196.03 & 402.06 & 425.75 \\ \hline
   $\sum$: & -9189.29 & 18462.58 & 18666.37 \\ 
   \hline
\end{tabular}
\caption{Log-likelihood, AIC$_F$ and BIC$_F$ of the fitted LGBN for the $cd3cd28+aktinhib$ experiment of the Sachs data on the original scale (F)}
\label{tab:GaussModX3}
\end{table}

%% file: 2-Dvineregression.tex
\section{Copulas and vine distributions}
\label{sec:copula}
For the analysis of large multivariate data sets multivariate statistical models are required, which can adequately describe not only the data center  but also their tail behavior. For this standard multivariate distributions such as the multivariate normal or Student $t$ distribution are insufficient. They often require that all univariate and multivariate marginal distributions are of the same type and might result in symmetric tail dependence of pairs of variables. These characteristics are often not satisfied and so the copula approach developed by \cite{Sklar} allows to separate the marginal behavior from the dependence structure. In particular
a $d$-dimensional copula $C$ is a multivariate
distribution function on the $d$-dimensional hyper cube $[0,1]^d$ with uniformly distributed marginals. 
The corresponding copula density for an  absolutely continuous copula we denote by $c$.
The fundamental theorem of \cite{Sklar} 
for  a $d$-dimensional random vector $\fX$ with joint distribution function $F$ and marginal distribution functions $F_j, j=1,\ldots,d$ 
is given by
\begin{align}
\label{sklar-cdf}
F(x_1,...,x_d) &= C(F_1(x_1),...,F_d(x_d))
\end{align} for some $d$-dimensional copula $C$. When $\bm X$ is absolutely continuous
the associated density $f$ can be expressed as
\begin{align}
\label{sklar-den}
f(x_1,...,x_d) &= c(F_1(x_1),...,F_d(x_d)) f_1(x_1)...f_d(x_d)
\end{align}
with marginal densities $f_j, j=\ldots, d$ and copula density $c$.

For absolutely continuous distributions the copula C is unique. Using \eqref{sklar-cdf} or \eqref{sklar-den} flexible multivariate distributions can be constructed if $d$-dimensional copulas are used. By inversion of \eqref{sklar-cdf} we can use any $d$-dimensional distribution function to obtain the corresponding copula. Examples for such copulas are the Gaussian and the Student $t$ copula. Note that using these copula families together with arbitrary margins results in multivariate distributions, sometimes also called meta distributions, which are much more flexible than the multivariate distribution classes used in the inversion. Another class of parametric copulas are built directly using generator functions is the class of Archimedean copulas. The Gumbel, Clayton and Frank copula families are prime examples. A nice introduction to copulas is given in \cite{genest2007} and more theoretical treatments are the books by \cite{nelsen2007introduction} and \cite{Joe1997}.

From \eqref{sklar-den} for $d=2$ we immediately can derive expression for the conditional density and distribution functions, which are needed later. In particular 
the conditional density $f_{1|2}$ and distribution function $F_{1|2}$ can be expressed as
\begin{align}
f_{1|2}(x_1|x_2) & = c_{1 2 }(F_1(x_1), F_2(x_2)) f_2(x_2) \label{cond-den}\\
F_{1|2}(x_1|x_2) 
& =\frac{\partial}{\partial F_2(x_2)}
 C_{1 2}(F_1(x_1),F_2(x_2)) \label{cond-dist}.
\end{align}

Since vine copulas are built out of bivariate copulas we give some properties of bivariate copulas. For this we use pairwise dependence measures such as Kendall's $\tau$ and Spearman's $\rho$, which are completely determined by the copula to measure the overall strength of the dependence. To assess the tail dependence upper and lower tail dependence coefficients can be used. 
It is easy to see that the Gaussian and the Frank copula do not have tail dependence, while the Student $t$ copula has symmetric tail dependence. Further the Clayton copula has no lower tail dependence while the Gumbel has upper tail dependence. For more details see for example Chapter 2 of \cite{czadobook}.

To allow for a visual comparison between different bivariate copula families marginally normalized contour plots are helpful. For this we consider three different scales: the original scale $(X_1,X_2)^\top$, the copula scale $(U_1, U_2)^\top = (F_1(X_1), F_2(X_2))^\top$ and the marginally normalized scale (z-scale) $(Z_1,Z_2)^\top=(\Phi^{-1}(U_1),\Phi^{-1}(U_2))^\top$. Here $\Phi$ denotes the distribution function of a $N(0,1)$ random variable. Comparison of contours on the copula scale for different families is difficult, since copula densities are in general unbounded at the corners of $[0,1]^2$. This is not the case if one works on the z-scale. Here $(Z_1,Z_2)^\top$ has $N(0,1)$ margins and thus any non-elliptical contour shape indicates a deviation from a Gaussian dependence. 

We now turn to estimation in a parametric setup. In a copula based model specified by \eqref{sklar-cdf} or \eqref{sklar-den} we have to estimate both the marginal and copula parameters. For this joint maximum likelihood estimation can be used, if the number of parameters are not too large. However it is more common to use a two step approach, by first estimating the marginal parameters based on i.i.d observations $\fx_i=(x_{i1},\ldots, x_{id})^\top$ for $i=1,\ldots, n$. This can be done separately for each of the $d$ margins. In a second step pseudo copula data is formed by setting
\begin{equation}
\label{eq:pseudo}
\fu_i=(u_{i1},\ldots,u_{id})^\top:=(\hat{F}_1(x_{i1}),\ldots,\hat{F}_d(x_{id}))^\top \mbox{ for } i=1,...,n
\end{equation}
using the estimated probability integral transforms (PIT) $\hat{F}_j, j=1,\ldots,d$. Then the copula parameters are estimated based on the pseudo copula data.
If parametric marginal models are used  we speak of an inference for margins approach (IFM) and 
if the empirical distribution is applied to the margins we have a semiparametric approach. The efficiency of the IFM approach has been investigated by \cite{joe2005}, while the semiparametric approach has been proposed by \cite{Genest}. 

In the case where parametric marginal and copula models do not fit the data, we can also use a nonparametric approach. For the margins kernel density based estimates of the distribution function can be used. For the bivariate copula case also many nonparametric estimators are available. Their efficiency have been investigated in \cite{nagler2014kernel}. The results showed that bivariate kernel density based estimation is preferred and this option is also available in
\texttt{rvinecopulib}  of \cite{rvinecopulib2019} to fit the more general class of vine distributions, which are build using bivariate copulas. We now give a short exposition of this class.


While the catalogue of bivariate parametric copula families is large, this is not the case for $d>2$. Therefore the aim of early research in vine copulas was to find a way to construct multivariate copulas using only bivariate copulas as building blocks. 
The appropriate tool to obtain such a construction is conditioning and 
 \cite{Joe96} gave the first
 pair copula construction
of a multivariate copula in terms of distribution
functions, while \cite{Bedford3,Bedford2} 
independently developed constructions  in terms of densities. Additionally they provided a general framework to identify all possible constructions.
We shortly illustrate this construction for $d=3$ starting with the recursive factorization 
\begin{equation}
\label{fac3}
f(x_1,x_2,x_3)=f_{3|12}(x_3|x_1,x_2) f_{2|1}(x_2|x_1)f_1(x_1)
\end{equation}
and treat each term separately. 
To determine $f_{3|12}(x_3|x_1,x_2)$ we consider the bivariate conditional density $f_{13|2}(x_1,x_3|x_2)$. This density has $F_{1|2}(x_1|x_2) (f_{1|2}(x_1|x_2))$ and $F_{3|2}(x_3|x_2) (f_{3|2}(x_3|x_2))$ as marginal distributions (densities) with associated conditional copula density  $c_{13;2}(\cdot,\cdot;x_2)$. More specifically $c_{13;2}(\cdot,\cdot;x_2)$ denotes the copula density associated with the conditional distribution of $(X_1,X_3)$ given $X_2=x_2$. Applying
 \eqref{sklar-den} to $f_{13|2}(x_1,x_3|x_2)$ gives
\begin{align}
\label{fac3.1-bi}
    f_{13|2}(x_1,x_3|x_2) & =  c_{13;2}(F_{1|2}(x_1|x_2),F_{3|2}(x_3|x_2);x_2) f_{1|2}(x_1|x_2)
    f_{3|2}(x_3|x_2).
\end{align}
Now $f_{3|12}(x_3|x_1,x_2)$ is the conditional density of $X_3$ given $X_1=x_1,X_2=x_2$ which can be determined using \eqref{cond-den}
applied to \eqref{fac3.1-bi} yielding
\begin{align}
\label{fac3.1}
f_{3|12}(x_3|x_1,x_2) & =
c_{13;2}(F_{1|2}(x_1|x_2),F_{3|2}(x_3|x_2);x_2) f_{3|2}(x_3|x_2).
\end{align}
Finally direct application of \eqref{cond-den}
gives 
\begin{align}
\label{fac3.2}
f_{2|1}(x_2|x_1) & = c_{12}(F_1(x_1),F_2(x_2)) f_2(x_2)\\
\label{fac3.3}
f_{3|2}(x_3|x_2) & = c_{23}(F_2(x_2),F_3(x_3))f_3(x_3).
\end{align}
Inserting \eqref{fac3.1}, \eqref{fac3.2} and \eqref{fac3.3} into \eqref{fac3} yields a pair copula decomposition
of an arbitrary three dimensional
density $f(x_1,x_2,x_3)$  as
\begin{align}
\label{pcc3}
f(x_1,x_2,x_3)  & =   c_{13;2}(F_{1|2}(x_1|x_2),F_{3|2}(x_3|x_2);x_2)
\times c_{23}(F_2(x_2),F_3(x_3))\\
& \times c_{12}(F_1(x_1),F_2(x_2)) f_3(x_3) f_2(x_2) f_1(x_1).
\nonumber
\end{align}
We see that the joint three dimensional density can be expressed in terms of bivariate copulas and (conditional) distribution
functions. However this decomposition is not unique, since
\begin{align}
\label{pcc3-1}
f(x_1,x_2,x_3) & =  c_{12;3}(F_{1|3}(x_1|x_3),F_{2|1}(x_2|x_1);x_3)
\times c_{13}(F_1(x_1),F_3(x_3))\\
& \times\ c_{23}(F_2(x_2),F_3(x_3)) f_3(x_3) f_2(x_2) f_1(x_1)
\nonumber \\
f(x_1,x_2,x_3) & =    c_{23;1}(F_{2|1}(x_2|x_1),F_{3|1}(x_3|x_1);x_1)
\times c_{13}(F_1(x_1),F_3(x_3))\\
& \times\ c_{12}(F_1(x_1),F_2(x_2)) f_3(x_3) f_2(x_2) f_1(x_1)
\nonumber
\end{align}
are two different decompositions using a reordering of the
variables in \eqref{fac3}. In these decompositions we have in general different conditional copula densities when the value of the conditioning variable is changing. This is intractable for estimation and therefore often the so called simplifying assumption is made, i.e the dependence on the conditioning value is ignored. For example we set $c_{13;2}(F_{1|2}(x_1|x_2),F_{3|2}(x_3|x_2);x_2)=c_{13;2}(F_{1|2}(x_1|x_2),F_{3|2}(x_3|x_2))$ in \eqref{pcc3} for every choice of $x_2$. Bivariate copulas are also referred to as pair copulas. Using the simplifying assumption in \eqref{pcc3} we speak of a pair copula construction and not of a decomposition. In this setup we can estimate both marginal and copula parameters in a parametric setup or use a nonparametric approach for both margins and bivariate copulas. Note that the specification of marginal and pair copula distributions can be done independently thus we have constructed a very flexible class of three dimensional densities. 

This construction principle involving only marginal distributions and pair copulas can be extended to arbitrary dimensions.
For arbitrary dimension we need a building plan  identifying which bivariate conditional distributions and their associated bivariate copulas are needed in the construction. This building plan is specified through a set of linked trees, where nodes in the next tree are edges of the current tree. Further the edges of the next tree have to satisfy the so called proximity condition. This tree structure was called by \cite{Bedford3,Bedford2} a vine tree structure. The corresponding edges identify the pair copulas needed to construct the $d$-dimensional density called a regular (R-) vine distribution. Since the number of such vine tree structures grows super exponentially \citep{morales2011count}, we need a greedy algorithm such as the one developed by \cite{dissmann2013selecting}. Under the simplifying assumption parameter estimation can be  performed in a stepwise fashion over the pair copula terms in a tree. This approach was developed by \cite{aas2009pair} and applied to financial data. There are two simple sub classes of R-vine distributions. The first one is called a D-vine distribution and uses as allowed tree structures only a path of all nodes in the tree. The class of C-vines occur in the case that all trees are stars with a root node in the center. 

For a detailed introduction to pair copula constructions and the resulting vine distributions consult \cite{czadobook} and \cite{joe2014dependence}. The use of such vine copula based models has grown tremendously. \cite{aas2016pair} and  
\cite{czadonagler2022} give reviews and point to current developments.

\section{Vine copula based regression models}
\label{sec:Vcbrm}
After constructing multivariate distributions we are interested in constructing a vine copula based regression model to allow for nonlinear regression effects. In a first approach
\cite{chang2019prediction, chang19a, cooke2019vine} build a vine distribution for the the covariates only and then add the response to the vine tree structure such that the resulting structure remains a vine tree structure and allows to specify the conditional response distribution in closed form. In this way numerical integration over the covariates is avoided. This first focus on the joint distribution of the covariates is not natural in the context of a regression problem, therefore an alternative is to start with the response as first node and add covariates one by one in such a way that again the resulting structure is a vine tree structure and the associated density of the response given the covariates is available in closed form. The adding a covariates stops if the conditional (penalized) log-likelihood  
does not increase any more when a further covariate is added. This approach was first developed
 for D-vine models by \citet{kraus2017d} and later extended to certain R-vine structures by \citet{zhu2021simplified}. \citet{tepegjozova2021nonparametric} considered both D- and C-vines, but generalized the procedure to look two steps ahead before adding a new covariate.
While the D- and C-vine based regression methods are feasible in large dimensions, the search for R-vines based regression is restricted to smaller dimensions because of the huge number of allowed R-vine tree structures. 

\section{D-vine copula based structural equation models}
\label{sec:Vcsem}
We now extend the LGBN model \eqref{eq:LGBN} to allow for non linear effects of the parents on a node. For this we recall the factorization \eqref{eq:decomp} and formulate for the conditional density of the node given its parents a vine copula based regression model. If we use the D-vine based regression formulation discussed in Section \ref{sec:Vcbrm} then we can express the conditional density of node $X_i$ given its parents $\bm\pi(X_i)=(X_{j_1(i)},\ldots, X_{j_{n_i}(i)})^\top$, where $n_i$ is the number of parents of node $X_i$ for $i=1,\ldots,d$ as

\begin{align}
\label{eq:dvinesem}
   f(x_i|\bm \pi(X_i)=\bm \pi(x_i))= &
  \left[ \prod_{s=2}^{n_i}
   c_{i,s;1:s-1}\right] \times c_{i,1}
   \times f_i(x_i),
  \end{align}
  where 
  $$c_{i,s;1:s-1}= c_{i,s;1:s-1}(F_{i|1:s-1}(x_i|x_{j_1(i)},\ldots x_{j_{s-1}(i)}),F_{s|1:s-1}(x_s|x_{j_1(i)},\ldots x_{j_{s-1}(i)}))$$
  and $c_{i,1}=c_{i,1}(F(x_i),F(x_{j_{1}(i)}))$.
  This will now form the building block for the factorization \eqref{eq:decomp} and in view of \eqref{eq:LGBN} we call the model specified by \eqref{eq:decomp} and \eqref{eq:dvinesem} a D-vine copula based SEM. 
  
  The order of the parent nodes 
  $\bm\pi(X_i)=(X_{j_1(i)},\ldots, X_{j_{n_i}(i)})^\top$ in the D-vine based conditional density \eqref{eq:dvinesem}
  is apriori not fixed, we will however later choose the order as obtained by using the methods developed by \cite{kraus2017d} and described in Section \ref{sec:Vcbrm}. Thus we start with the most important parent node with regard to the conditional (penalized) log-likelihood of the node and continue adding parent nodes until this quantity can no longer improved or all parent nodes are included. 
  
  Since the forward selection of the parent nodes also includes the possibility of not including all parents this gives a way to remove edges from  the starting consent graph. These edges are then not supported by the data. We will now apply this class of copula based SEMs to the
  experimental setting $cd3cd28+aktinhib$ of the Sachs data.

%% file: 4-vine-analysis.tex
\section{Analysis of experimental setting 
\texorpdfstring{$cd3cd28+aktinhib$}{cd3cd28+aktinhib}
 from the Sachs data}
 For this analysis we used the data from
 \url{ https://science.sciencemag.org/content/308/5721/523/tab-figures-data} on the logarithmic scale.
Recall that using the consent graph given in Figure \ref{fig:DAG} results in the density decomposition \eqref{eq:LGBNDicht} for the eleven protein variables.
Next we want to utilize the D-vine copula based SEM introduced in Section \ref{sec:Vcsem}. This entails that every density term in \eqref{eq:LGBNDicht} is modeled by using the D-vine regression introduced in Section
\ref{sec:Vcbrm}. 
Using \eqref{eq:dvinesem} it follows that  (\ref{eq:LGBNDicht}) can be rewritten as

\begin{align}
\label{eq:SachsDV}
f(\bm x)
=&\ c_{(akt, 1)}*c_{(akt, 2;\ 1)}*c_{(akt , 3;\ 1,\ 2)}*f(akt)*c_{(erk, 1)}*c_{(erk,2;\ 1)}*f(erk)\nonumber \\
&*c_{(mek,1)}*c_{(mek, 2;\ 1)}*c_{(mek, 3;\ 1,\ 2)}*f(mek)\\\nonumber
&*c_{(raf, 1)}*c_{(raf, 2;\ 1)}*f(raf)*c_{(jnk, 1)}*c_{(jnk, 2;\ 1)}*f(jnk)\\\nonumber
&*c_{(p38\ 1)}*c_{(p38, 2;\ 1)}*f(p38)*c_{(pka, pkc)}*f(pka)\\\nonumber
&*c_{(pkc, 1)}*c_{(pkc, 2;\ 1)}*f(pkc)*c_{(pip2\ 1)}*c_{(pip2, 2;\ 1)}*f(pip2)\\\nonumber
&*c_{(plc, pip3)}*f(plc)*f(pip3).\nonumber
\end{align}

Note that if a node has more than one parent, the order of the D-vines might change depending on the chosen marginals and copulas. In this case we use numbers to denote the position in the order of the D-vine instead of the explicit nodes. Hence, "$1$" denotes the first parent of the modeled node, "$2$" the second parent in the D-vine regression and so on.
For our analysis we have 845 observations available for the $cd3cd28+aktinhib$ experiment. In this experiment the reagent
Anti-CD3/CD28 \footnote[1]{General perturbation: Activates T cells and induces proliferation and cytokine
production. Induced signaling through the T cell receptor (TCR), activated ZAP70,
Lck, PLC-g, Raf, Mek, Erk, and PKC. The TCR signaling converges on transcription
factors NFkB, NFAT, and AP-1 to initiate IL-2 transcription \citep{Sachs2005}} and 
AKT \footnote[2]{Inhibitor specific perturbation: Binds inositol pleckstrin domain of AKT and blocks AKT
translocation to the membrane where normally AKT becomes phosphorylated and
active [median inhibitory concentration (IC50) 0 5 mM]. Inhibition of AKT and
phosphorylation of AKT substrates are needed to enhance cell survival \citep{Sachs2005}}  are  having an activation effect on the protein $pip3$ and $pip2$ and an inhibitor effect on $akt$ according to Figure 2 of \cite{Sachs2005}.

To estimate the parameters  of the marginal and copula terms in Equation (\ref{eq:SachsDV})  we follow the IFM approach discussed in Section \ref{sec:copula}. So in a first step we have to estimate the margins and then form appropriate pseudo data as in \eqref{eq:pseudo}. In the second step we fit a D-vine copula to the node as first variable and its parents as remaining variables in the described forward  approach.

To see how the fit of the model varies, when using different marginals (\textbf{M}) or copulas (\textbf{C}) we fit three different copula based SEMs. For the first model we use kernel density estimates as margins and allow for parametric and non-parametric copulas (\textbf{M}$^{\mathbf{ker}}$\textbf{C}$^{\mathbf{pnp}}$), for the second one we use Gaussian mixture margins  since we observe some multi nodal margins  and the same set of copulas (\textbf{M}$^{\mathbf{par}}$\textbf{C}$^{\mathbf{pnp}}$) and for the last model  we only allow for Gaussian margins and Gaussian copulas (\textbf{M}$^{\mathbf{gauss}}$\textbf{C}$^{\mathbf{gauss}}$). A non parametric pair copula is chosen, when it fits better than any implemented parametric family and is implemented in \texttt{vinereg} of \cite{vinereg}.

The fit of the different marginals is shown in Figure \ref{fig:contALL3X3}
of Appendix A1
comparing marginal histograms to the fitted density, qq plots and checking if the data after applying the PIT using the estimated marginal densities is approximately uniformly distributed.
We observe that using Gaussian margins for the PIT in general does not result in uniformly distributed data indicating a bad fit. On the other hand, Gaussian mixture margins and kernel density estimates  fit the margins very well. It is hard to distinguish between these two models. 
We also consider the goodness of fit measures log-likelihood, AIC$_M$ and BIC$_M$ of the fitted marginals. The results are displayed in Table \ref{tab:Margins} in Appendix A2 
and show that Gaussian margins give the worst fit. 


To support the use of copula based SEMs instead of linear Gaussian ones, we give normalized contour plots in Figure \ref{fig:pairsPlotsMarg} of Appendix A1. 
From there we see non elliptical shapes in the lower triangular panels regardless which marginal models were used, thus pointing towards the need to capture non Gaussian dependence. 


In the next step we use the fitted marginals to transform the data to the copula scale by applying the estimated PIT. On these new data sets we then fit copulas according to three defined models. 

Using kernel density estimates for the PIT and then fit a D-vine regression model for each node with at least one parent results in a log-likelihood of \textbf{2316.23}, an AIC$_C$ of \textbf{-4228.73} and a BIC$_C$ of \textbf{-3272.06} of the copula terms. Using Gaussian mixture margins and the same set of copula families results in a log-likelihood of \textbf{2292.34}, an AIC$_C$ of \textbf{-4189.49} and a BIC$_C$ of \textbf{-3252.98} of the copula terms. The last model where Gaussian margins and Gaussian copulas are used worsens the three measures to a log-likelihood of \textbf{1497.84}, an AIC$_C$ of \textbf{-2971.68} and a BIC$_C$ of \textbf{-2914.80} of the copula terms. More detailed results are displayed in Table \ref{tab:Cops} and Table \ref{tab:CallMkerCopX3} on the copula scale as well as contour plots of the fitted copulas in Figure \ref{fig:contX3c} of Appendix A3. They show that Gaussian pair copulas are only seldom chosen in the non Gaussian SEMs. Further the parametric pair copula families contained in \texttt{vinereg} are not sufficient and many nonparametric pair copulas (\texttt{tll}) are selected.
We see strong similarities on the fitted copula scale  between the \textbf{M}$^{\mathbf{ker}}$\textbf{C}$^{\mathbf{pnp}}$ model and the \textbf{M}$^{\mathbf{mix}}$\textbf{C}$^{\mathbf{pnp}}$ model. 

If a parent $j(i)$ is not selected in the D-vine regression for the node $i$, this indicates that the copula density $c_{i,j(i);1:j(i)-1}$ is very close to the bivariate independence copula density 
($c_{ind}(u_1,u_2)=1$ for every $(u_1,u_2) \in [0,1]^2$) and thus the edge $j(i)\rightarrow i$ is not supported by the data.
This implies that in both copula based non Gaussian SEMs only 17 of the 20 dependencies given by the biological consent graph are present for this experimental setting (see Panel (b) of Figure \ref{fig:DAG}). The removed edges are $mek \rightarrow erk, plc \rightarrow pkc$ and $pip3 \rightarrow akt$. The deletion of $mek \rightarrow erk$ implies
that there is no pathway from $erk$ to $akt$. This might be a consequence of the inhibiting effect on $akt$. Further the deletion of $plc \rightarrow pkc$ implies that $pkc$ can only be reached from $plc$ over $pip2$, which is sensible since in this experiment $pip2$ is activated. The removal of $pip3 \rightarrow akt$ is also sensible since $akt$ is inhibited in this setting. The edges $plc \rightarrow pkc$ and $pip3 \rightarrow akt$ also were identified as edges to be removed from the consent graph in the original analysis of \cite{Sachs2005}.

The number of supported edges decreases for the \textbf{M}$^{\mathbf{gauss}}$\textbf{C}$^{\mathbf{gauss}}$ model further. Here only 12 edges are modelled. The removal of these additional four edges (colored red in Panel (b) of Figure \ref{fig:DAG}) is questionable, since as we show now that the fit of the \textbf{M}$^{\mathbf{gauss}}$\textbf{C}$^{\mathbf{gauss}}$ is inferior to the other two copula based SEM's.

We now include the marginal fit for the different SEMs based on Equation \eqref{eq:dvinesem} to compute overall fit measures, which allow us to also compare to the performance of the standard LGBN. These are contained in Table \ref{tab:sumMod}.
\begin{table}
  \centering
  \tiny
  \begin{tabular}{|r|r|r|r|r|r|r|r|r|r|r|r||r|}
 \hline 
Model & raf & mek & plc & pip2 & pip3 & erk & akt & pka & pkc & p38 & jnk & $\sum$: \\
 \hline \hline
LGBN & -970 & -552 & -922 & -1249 & -987 & -968 & -196 & -831 & -952 & -571 & -992 & -9189 \\   \hline

\textbf{M}$^{\mathbf{ker}}$\textbf{C}$^{\mathbf{pnp}}$ & -902 & -399 & -827 & -921 & -895 & -822 & -79 & -766 & -770 & -246 & -837 & -7464 \\   \hline

\textbf{M}$^{\mathbf{mix}}$\textbf{C}$^{\mathbf{pnp}}$ & -907 & -416 & -826 & -934 & -917 & -830 & -103 & -768 & -775 & -263 & -845 & -7587 \\ \hline

\textbf{M}$^{\mathbf{gauss}}$\textbf{C}$^{\mathbf{gauss}}$& -970 & -552 & -922 & -1249 & -987 & -968 & -197 & -831 & -952 & -571 & -993 & -9193 \\ 
   \hline
\end{tabular}

\caption*{(a) Log-likelihood of the conditional densities}
\bigskip

  \begin{tabular}{|r|r|r|r|r|r|r|r|r|r|r|r||r|}
 \hline 
Model & raf & mek & plc & pip2 & pip3 & erk & akt & pka & pkc & p38 & jnk & $\sum$: \\
 \hline \hline
LGBN & 1948 & 1113 & 1850 & 2507 & 1978 & 1944 & 402 & 1668 & 1913 & 1149 & 1992 & 18463 \\   \hline

\textbf{M}$^{\mathbf{ker}}$\textbf{C}$^{\mathbf{pnp}}$ & 1858 & 876 & 1718 & 1938 & 1813 & 1695 & 236 & 1552 & 1556 & 565 & 1750 & 15558 \\   \hline

\textbf{M}$^{\mathbf{mix}}$\textbf{C}$^{\mathbf{pnp}}$ & 1854 & 889 & 1710 & 1967 & 1844 & 1702 & 276 & 1555 & 1569 & 590 & 1747 & 15703 \\ \hline

\textbf{M}$^{\mathbf{gauss}}$\textbf{C}$^{\mathbf{gauss}}$& 1948 & 1113 & 1850 & 2505 & 1978 & 1943 & 401 & 1666 & 1911 & 1148 & 1991 & 18453 \\
   \hline
\end{tabular}

\caption*{(b) AIC$_F$ of the conditional densities}
\bigskip

  \begin{tabular}{|r|r|r|r|r|r|r|r|r|r|r|r||r|}
 \hline 
Model & raf & mek & plc & pip2 & pip3 & erk & akt & pka & pkc & p38 & jnk & $\sum$: \\
 \hline \hline
LGBN & 1967 & 1137 & 1864 & 2526 & 1987 & 1963 & 426 & 1687 & 1932 & 1168 & 2011 & 18666 \\    \hline

\textbf{M}$^{\mathbf{ker}}$\textbf{C}$^{\mathbf{pnp}}$ & 1989 & 1057 & 1870 & 2165 & 1868 & 1817 & 421 & 1599 & 1596 & 738 & 1930 & 17051 \\    \hline

\textbf{M}$^{\mathbf{mix}}$\textbf{C}$^{\mathbf{pnp}}$ & 1946 & 1021 & 1847 & 2202 & 1867 & 1800 & 441 & 1597 & 1611 & 740 & 1883 & 16957 \\ \hline

\textbf{M}$^{\mathbf{gauss}}$\textbf{C}$^{\mathbf{gauss}}$& 1967 & 1132 & 1864 & 2519 & 1987 & 1957 & 420 & 1676 & 1925 & 1162 & 2005 & 18614 \\  
   \hline
\end{tabular}

\caption*{(c) BIC$_F$ of the conditional densities}
\bigskip

  \begin{tabular}{|r|r|r|r|r|r|r|r|r|r|r|r||r|}
 \hline 
Model & raf & mek & plc & pip2 & pip3 & erk & akt & pka & pkc & p38 & jnk & $\sum$: \\
 \hline \hline
LGBN & 4 & 5 & 3 & 4 & 2 & 4 & 5 & 3 & 4 & 4 & 4 & 42 \\  \hline

\textbf{M}$^{\mathbf{ker}}$\textbf{C}$^{\mathbf{pnp}}$ & 28 & 38 & 32 & 48 & 12 & 26 & 39 & 10 & 8 & 37 & 38 & 315 \\    \hline

\textbf{M}$^{\mathbf{mix}}$\textbf{C}$^{\mathbf{pnp}}$ & 20 & 28 & 29 & 49 & 5 & 21 & 35 & 9 & 9 & 32 & 29 & 265 \\  \hline

\textbf{M}$^{\mathbf{gauss}}$\textbf{C}$^{\mathbf{gauss}}$& 4 & 4 & 3 & 3 & 2 & 3 & 4 & 2 & 3 & 3 & 3 & 34 \\ 
   \hline
\end{tabular}
\caption*{(d) Number of (effective) parameters of the conditional densities}
\bigskip

\caption{Goodness of fit measures of the fitted copula based SEMs and LGBN}
 \label{tab:sumMod}
\end{table}

%% file: 5-validation.tex
We observe that the LGBN  and the  D-vine based SEM with Gaussian margins and copulas result in almost identical results for all three goodness of fit measures. This is reasonable as both of them specify a multivariate Gaussian distribution following \cite{KollerFriedman2009} and \cite{MORALES2008699}. The parameters of the conditional Gaussian distribution given by each node in the two models are displayed in Table \ref{tab:LGBNCOND} in the Supplement S1.
Many similarities between these two models are visible even though all possible dependencies are modelled in the LGBN, as it was fitted optimizing the conditional  log-likelihood with all parents of a node, whereas this is not the case for the \textbf{M}$^{\mathbf{gauss}}$\textbf{C}$^{\mathbf{gauss}}$. 

Comparing the two models to the \textbf{M}$^{\mathbf{ker}}$\textbf{C}$^{\mathbf{pnp}}$ model and the \textbf{M}$^{\mathbf{mix}}$\textbf{C}$^{\mathbf{pnp}}$ model we can see a strong improvement in all goodness of fit measures in the models with parametric and non-parametric copulas. This is no surprise since we have already seen that Gaussian mixture margins and kernel density estimates provide a much better fit than Gaussian margins and allowing flexibility in the choice of the pair copula families provides an advantage. We observe that between the non Gaussian models only slight differences are visible mostly due to the large mumber of effective parameters needed for the kernel density estimates, which influences the BIC$_F$.

\section{Model validation}
Up to this point we have only compared the goodness of fit measures between the four model but have not evaluated if they really reproduce structures visible in  the observed data. For this we sampled 845 times from each model by starting in the node $pip3$ and sampling it using its fitted marginal density. We then follow the topological order, i.e  for each variable in this order we simulate data for the node using the already simulated data for its parents. For the LGBN we use that we can express the conditional distribution as a conditional normal distribution. To sample from the D-vine copula based SEMs we utilize the algorithm presented in \cite{bevacqua2017multivariate}. 

  \begin{figure}[ht]
    \centering
    \subfloat[\textbf{M}$^{\mathbf{ker}}$\textbf{C}$^{\mathbf{pnp}}$]{{\includegraphics[scale=0.2]{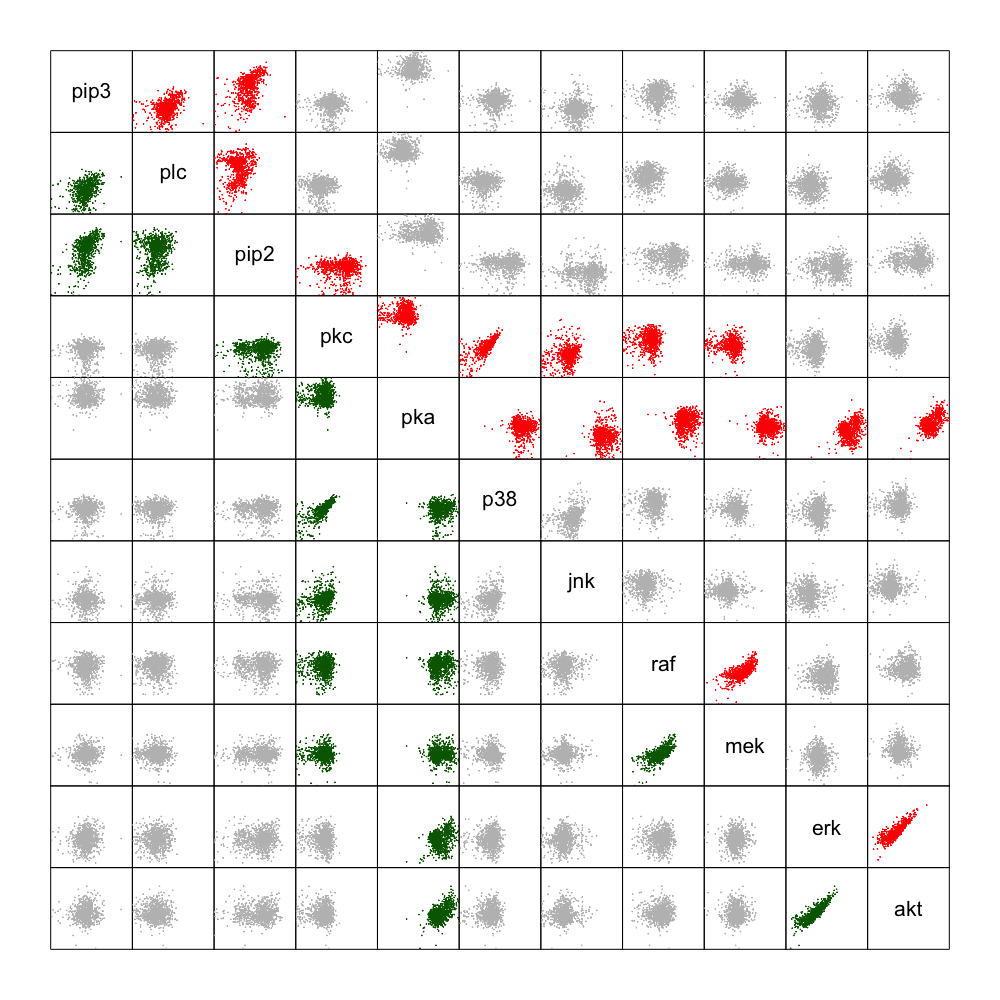}}}%
    \subfloat[\textbf{M}$^{\mathbf{mix}}$\textbf{C}$^{\mathbf{pnp}}$]{{\includegraphics[scale=0.2]{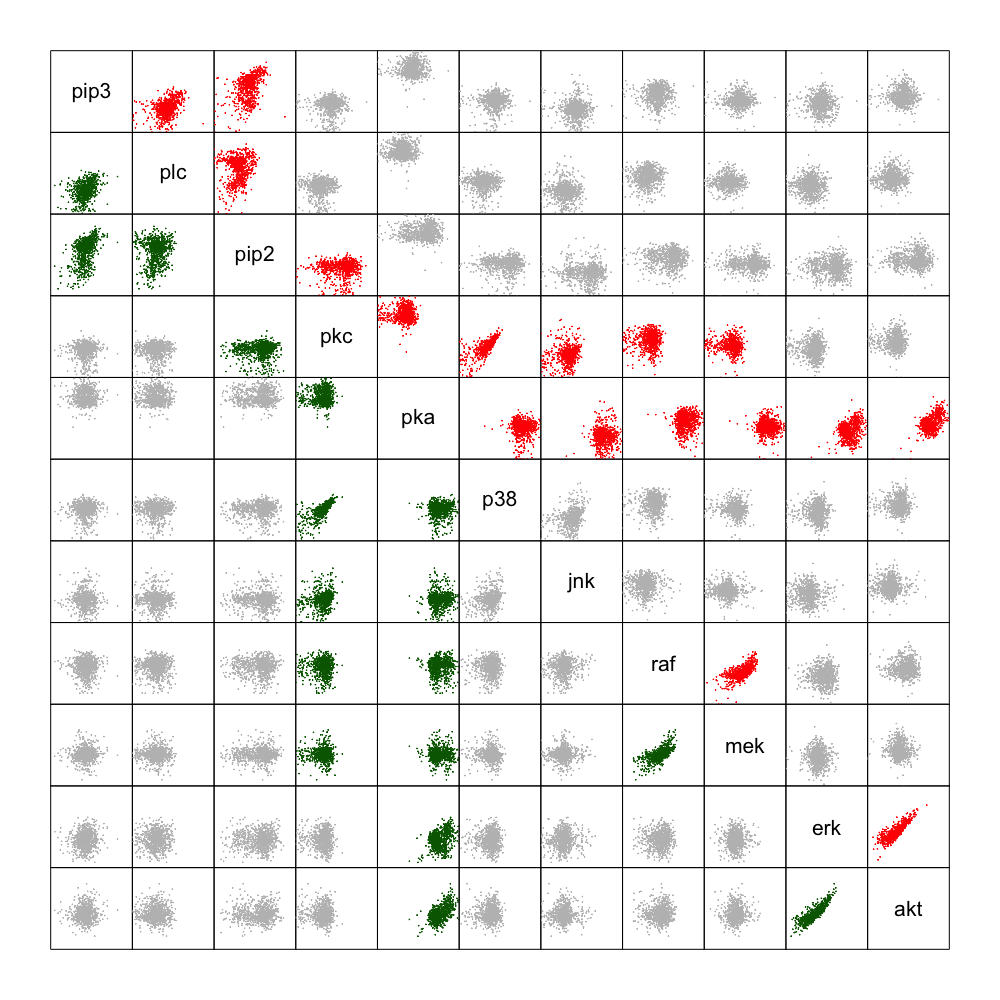}}}
    \bigskip
    
    \subfloat[\textbf{M}$^{\mathbf{gauss}}$\textbf{C}$^{\mathbf{gauss}}$]{{\includegraphics[scale=0.2]{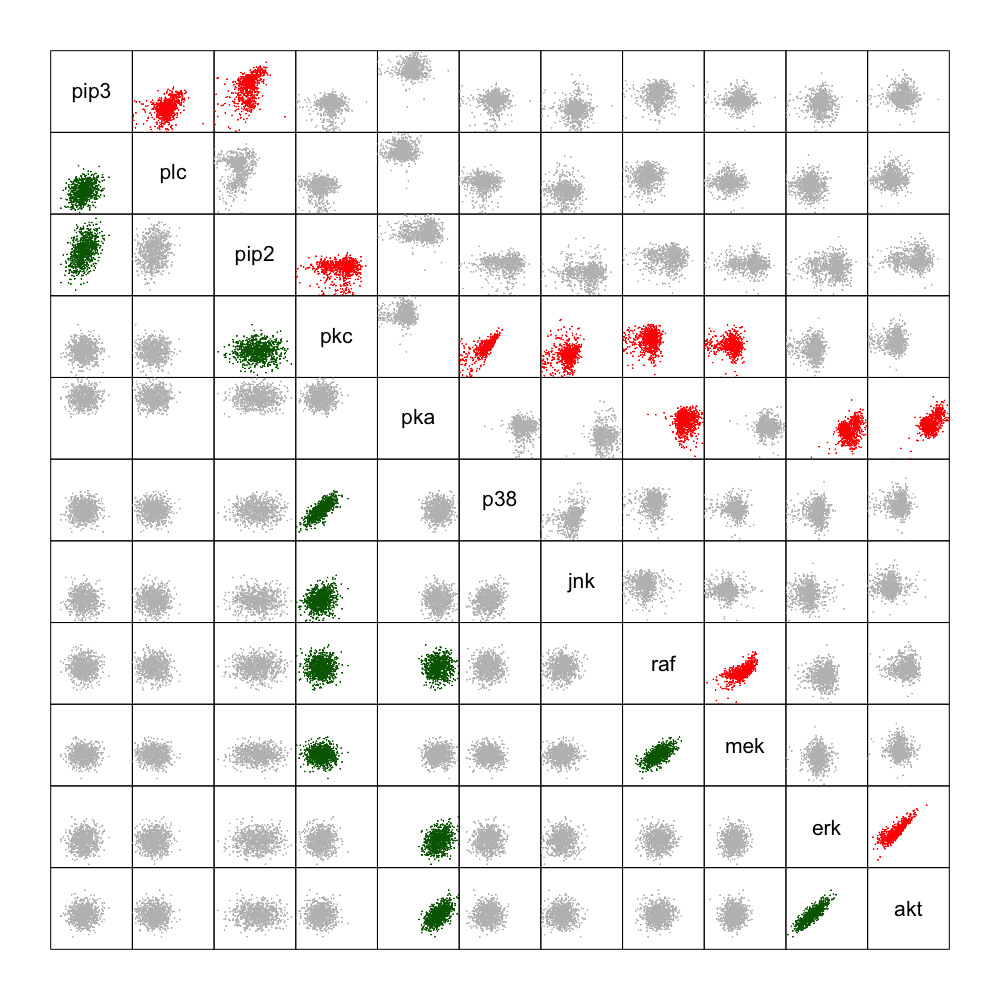}}}%
    \subfloat[LGBN]{{\includegraphics[scale=0.2]{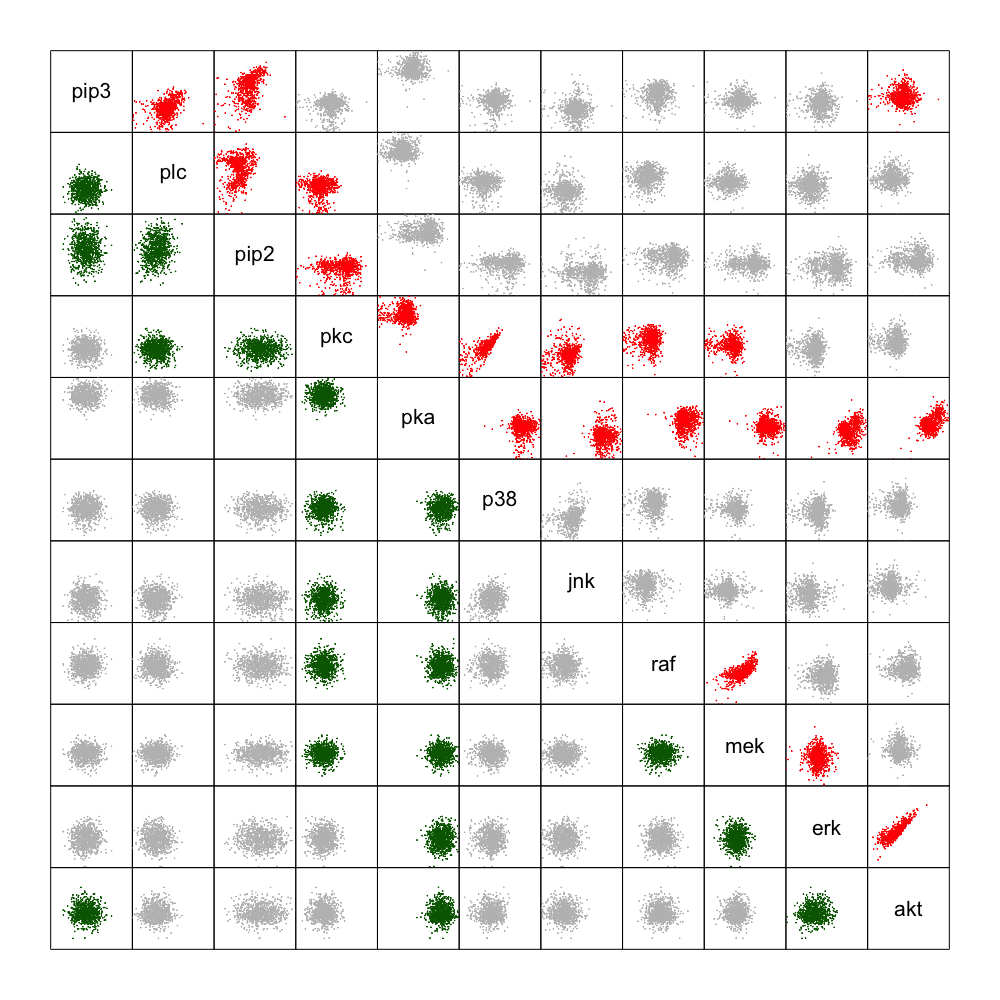}}}
    \caption{Scatter plots for each pair of nodes of the respective model in the lower panel and of the $cd3cd28+aktinhib$ data set in the upper panel. Green plots in the lower panel indicate that an edge exists between these two nodes in the model. For a better visibility the same pairs are colored in red in the upper panel}%
    \label{fig:contX3}
\end{figure}

To compare the induced pairwise dependence structures we consider the scatter plots  for node pairs of the observed $cd3cd28+aktinhib$ data and of the simulated data sets from the different models in Figure \ref{fig:contX3}. Looking at the scatter plots of the \textbf{M}$^{\mathbf{ker}}$\textbf{C}$^{\mathbf{pnp}}$ model and the \textbf{M}$^{\mathbf{mix}}$\textbf{C}$^{\mathbf{pnp}}$ model we observe almost no differences. This does not surprise as we have already seen that their fitted copulas are almost identical. Comparing the scatter plots to the observed scatter from the $cd3cd28+aktinhib$ data set we see that they closely match.
 For all variable pairs the shapes, even more complicated ones like between the node pairs $(raf, mek)$ or $(erk, akt)$, are similar from the simulated data of the two fitted non Gaussian SEMs and the original data.

\begin{figure}[ht]
\centering
\includegraphics[width=0.66\textwidth]{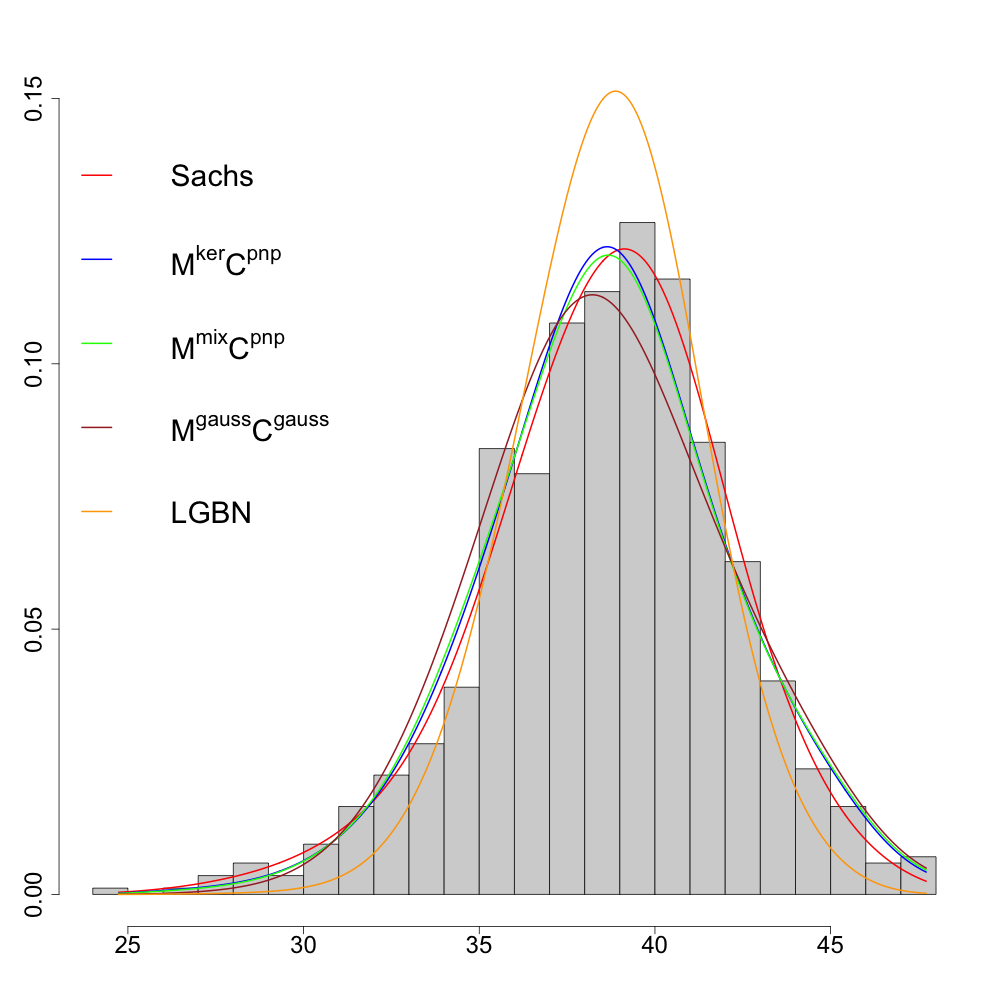}
\caption{Histograms of the sum over all nodes in the $cd3cd28+aktinhib$ experiment and fitted kernel density estimates to the simulated data sets of the different models}
\label{fig:kdesum}
\end{figure}

However comparing the scatter plots of the \textbf{M}$^{\mathbf{gauss}}$\textbf{C}$^{\mathbf{gauss}}$ model and the LGBN instead, it is clear that both of them have problems modeling more complicated shapes as they are only able to model elliptical shaped dependencies. It seems that the \textbf{M}$^{\mathbf{gauss}}$\textbf{C}$^{\mathbf{gauss}}$ model does this a little bit better looking at the pairs plots of $(raf,mek)$ and $(erk,akt)$ which are closer to the ones from the original data set than the ones from the LGBN.
\clearpage
Since in Figure \ref{fig:contX3} we only considered bivariate dependence properties, we  check if high/low values in several variables at once appear in a similar frequency in the $cd3cd28+aktinhib$ data and the simulated data. Hence, we sum up over the variables in each sample and then fit kernel density estimates to the resulting data. We can then compare the kernel density estimates fitted to the sum over the $cd3cd28+aktinhib$ data and simulated data from the investigated models. The results are displayed in Figure \ref{fig:kdesum}.

Comparing the fitted kernel density estimates of the sum over all nodes we see that especially the ones fitted to the simulated data from the \textbf{M}$^{\mathbf{ker}}$\textbf{C}$^{\mathbf{pnp}}$ model and the \textbf{M}$^{\mathbf{mix}}$\textbf{C}$^{\mathbf{pnp}}$ model are very close to the original data set and are only slightly shifted to the left. On the other hand, we observe that the ones fitted to the sum of the simulated data from the \textbf{M}$^{\mathbf{gauss}}$\textbf{C}$^{\mathbf{gauss}}$ model does not reach the height of the mode compared to the original data set whereas the LGBN overestimates the height of the mode. Hence, considering the ability of the different models to recreate the data set they are fitted on the \textbf{M}$^{\mathbf{ker}}$\textbf{C}$^{\mathbf{pnp}}$ model and the \textbf{M}$^{\mathbf{mix}}$\textbf{C}$^{\mathbf{pnp}}$ model outperform the other two.
\clearpage
\section{Conditional density and medians for given values of parent nodes}
To see the effect of allowing a more general dependence structure in the SEM \eqref{eq:dvinesem} compared to a LGBN we investigate the behavior of the conditional density of a node given its parents for the studied models. Here we like to investigate the effect when median and extreme values of the parents are used.
If a node only has one parent in the DAG, we chose the $10\%$, $50\%$ and $90\%$ empirical quantile  of the parent node as conditioning values. This is appropriate for the nodes $plc$ and $pka$ and the associated conditioning values are shown  in Figure \ref{fig:1dim} of Supplement S2. 

Since there is no simple equivalent of quantiles in two or three dimensional distributions a different approach had to be used for all other nodes. 
If a node has two or three parents, we fitted a two/three-dimensional kernel density estimate to the joint distribution of the parent nodes using the $cd3cd28+aktinhib$ data. We then manually choose one point close to the mode of the fitted density, two in the tails and two more which are neither close to the mode nor in the tails. This is illustrated in Figure \ref{fig:2dim} for nodes with two parents and Figure \ref{fig:3dim} for nodes with three parents, respectively, in Supplement S2. 
Note that while six nodes have two parents, ($raf$, $pip2$, $erk$, $pkc$, $p38$ , $jnk$), three of them 
($raf$, $p38$ and $jnk$), have the same set of parents, namely $pka$ and $pkc$, so we only obtain four cases for the values of the conditioning variables. The chosen values of the parent nodes are summarized 
in Table \ref{tab:Cond}.
\begin{table}[ht]
\centering
\begin{tabular}{|r|r|r|r|r|r|r|}
  \hline
  Node & Parent & Mode & \multicolumn{2}{c|}{Middle} & \multicolumn{2}{c|}{Tail}  \\
  \hline
  plc & pip3 & 2.68 &  & & 3.56 & 4.38   \\  \hline
  \multirow{2}{*}{pip2} & plc & 2.40 & 1.80 & 2.80 & 3.80 & 2.90 \\ 
    & pip3 & 3.50 & 3.60 & 3.90 & 4.30 & 2.10 \\ \hline
\multirow{2}{*}{pkc} & plc & 2.50 & 1.90 & 2.90 & 1.90 & 4.00 \\ 
    & pip2 & 4.90 & 5.20 & 4.80 & 2.50 & 5.00 \\  \hline
  pka & pkc & 1.80 &  &  & 2.93 & 3.46  \\ \hline
 \multirow{2}{*}{p38, jnk, raf} & pka& 6.00 & 6.00 & 7.00 & 7.50 & 6.50 \\ 
    & pkc & 3.00 & 2.50 & 3.00 & 2.50 & 1.75 \\  \hline
   \multirow{3}{*}{mek} & raf & 3.62 & 4.79 & 3.49 & 4.43 & 1.97 \\
    & pka & 6.10 & 6.52 & 7.40 & 7.64 & 6.63 \\ 
    & pkc & 2.88 & 3.44 & 3.10 & 2.04 & 2.71 \\  \hline
   \multirow{2}{*}{erk} & mek & 3.10 & 3.50 & 2.90 & 4.50 & 3.00 \\  
    & pka & 6.10 & 6.20 & 6.40 & 6.00 & 7.50 \\  \hline
   \multirow{3}{*}{akt} & erk & 2.57 & 4.22 & 2.10 & 2.58 & 1.54 \\  
    & pka & 6.19 & 7.15 & 5.89 & 6.28 & 6.63 \\
    & pip3 & 3.73 & 3.68 & 3.99 & 5.41 & 4.66 \\ \hline 
\end{tabular}
\caption{Chosen values of the parent nodes used for the density of a node given its parents}
\label{tab:Cond}
\end{table}

In the following we will denote the points which we chose close to the the mode of the fitted parent distribution as "Mode", the points in the tails as "Tail" and the points between them as "Middle". If a node only has one parent, we will assign the points from the $10\%$ and the $90\%$ quantile to "Tail" and the point from the $50\%$ quantile to "Mode". Note that as in the \textbf{M}$^{\mathbf{gauss}}$\textbf{C}$^{\mathbf{gauss}}$ model not all dependencies have been modeled, in this model sometimes the number of parents change with the model.

\begin{figure}[ht]
  \hspace{1cm} Mode  \hspace{2.2cm} Middle \hspace{3.5cm} Tail
    \centering
    \includegraphics[scale=0.1]{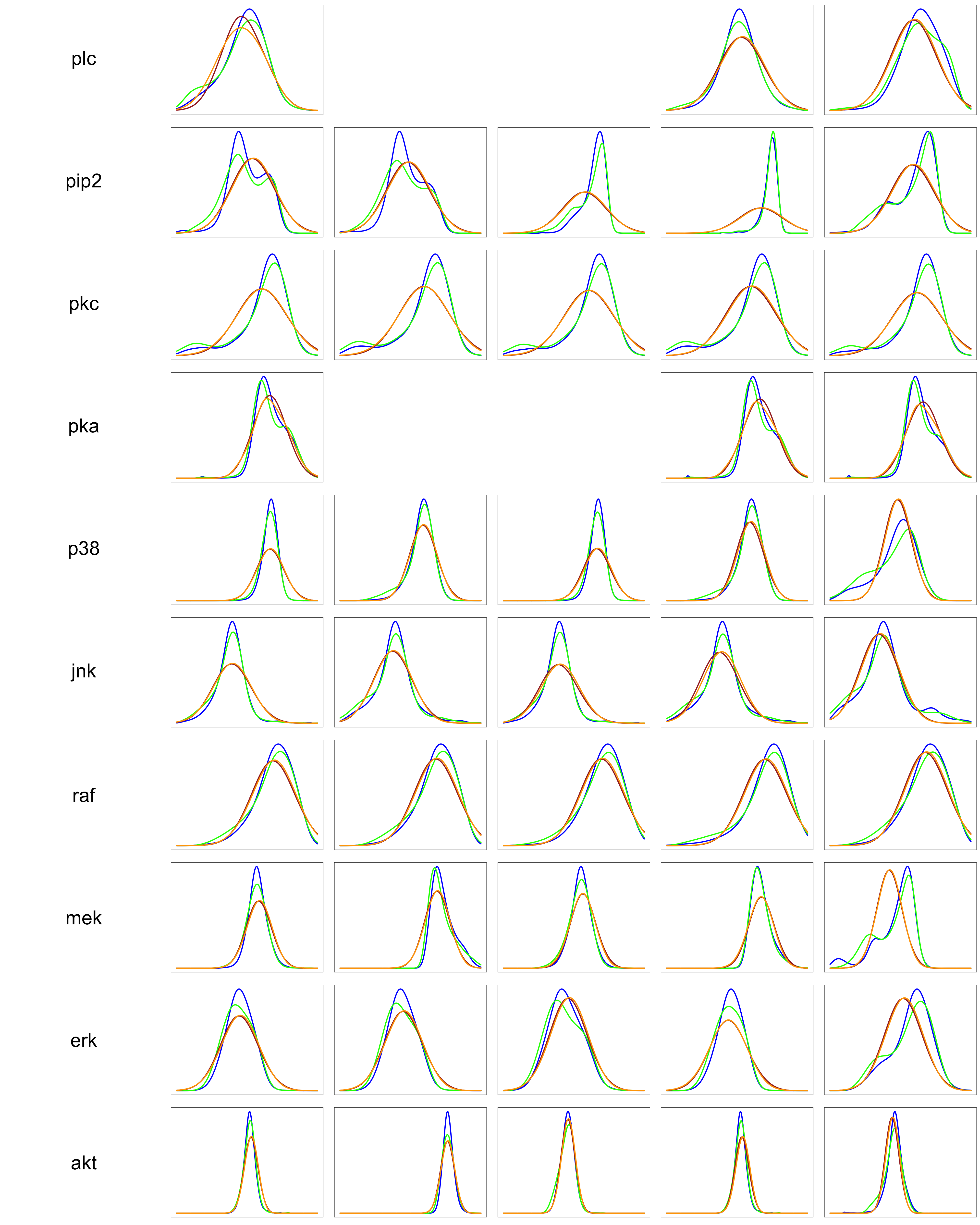}
    \caption{Density plots of the kernel density estimates fitted to the simulated data based on different structural equation models conditioned on values for the parent nodes in Table \ref{tab:Cond} following the order there. Blue:~\textbf{M}$^{\mathbf{ker}}$\textbf{C}$^{\mathbf{pnp}}$. Green:~\textbf{M}$^{\mathbf{mix}}$\textbf{C}$^{\mathbf{pnp}}$. Brown: \textbf{M}$^{\mathbf{gauss}}$\textbf{C}$^{\mathbf{gauss}}$. Orange: LGBN}%
    \label{fig:BedDicht}
\end{figure}

As earlier we sample 845 times from each node of the model given the values of the parent nodes and then fitted kernel density estimates to this sample. The resulting density plots are displayed in Figure \ref{fig:BedDicht}. As we expect, we observe only small differences in the general shape of the conditional density plots between  \textbf{M}$^{\mathbf{ker}}$\textbf{C}$^{\mathbf{pnp}}$ and \textbf{M}$^{\mathbf{mix}}$\textbf{C}$^{\mathbf{pnp}}$. The same holds when comparing the \textbf{M}$^{\mathbf{Gauss}}$\textbf{C}$^{\mathbf{Gauss}}$ model and the LGBN. However the shapes of the conditional density can be very different for the non Gaussian copula based SEMs and the Gaussian SEMs, thus illustrating very different conditional effects for the chosen parent values.


\clearpage

\begin{figure}[ht]
\centering
    \centering
    \subfloat[$\alpha=.10$ ]{{\includegraphics[width=0.29\textwidth]{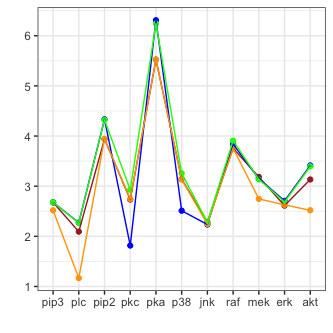}}}
    \qquad
    \subfloat[$\alpha=.50$ ]{{\includegraphics[width=0.29\textwidth]{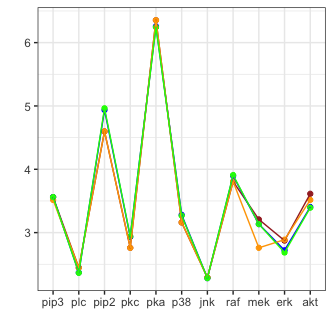}}}
    \qquad
        \subfloat[$\alpha=.90$]{{\includegraphics[width=0.29\textwidth]{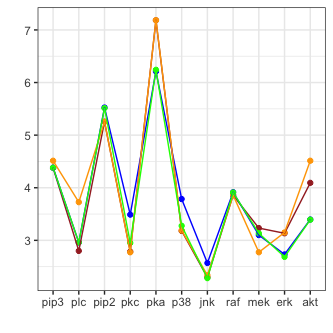}}}%
\caption{Conditional median of a node given the fitted median value of its parents starting with $pip3$ at the sample quantile at level $\alpha$ (Brown: \textbf{M}$^{\mathbf{gauss}}$\textbf{C}$^{\mathbf{gauss}}$, blue: \textbf{M}$^{\mathbf{ker}}$\textbf{C}$^{\mathbf{pnp}}$, green: \textbf{M}$^{\mathbf{mix}}$\textbf{C}$^{\mathbf{pnp}}$, orange: LGBN)}
\label{fig:DAG1}
\end{figure}

As a final visualisation of the differences on relevant conditional quantities we consider the conditional median for each node given its parents. 
We follow the topological order  as given in Table \ref{tab:TopOrd}. For $pip3$ we use a sample quantile of level $\alpha$ denoted by $q_\alpha(pip3)$ as value of pip3. Next we determine the median of plc given $pip3=q_\alpha(pip3)$ using the D-vine quantile regression of \cite{kraus2017d}, we denote this median as $med_\alpha(plc)$. For the node $pip2$ we determine the conditional median given $plc=med_\alpha(plc),pip3=q_\alpha(pip_3)$
using D-vine quantile regression. We denote this conditional median by $med_\alpha(pip2)$. We continue in this way until all conditional medians are determined. The needed parents for the different models can be found in Table A3 of Appendix A3. We investigate three different $\alpha=.10, .50 $ and $.90$ levels and the results are visualized in Figure \ref{fig:DAG1}. We see pronounced differences for the nodes $plc,pkc, mek,erk$ and $akt$.

%% file: 6-future-directions.tex
\section{Summary and future research directions}
In this paper we suggested to use a D-vine copula based SEM to analyze data which obeys a graphical structure. At least a starting graph has be known, but the data does not need to follow a joint Gaussian distribution. The appropriate copula regression model for the conditional distribution of the node given its parents is very flexible, since the bivariate building blocks of the D-vine regression can be independently chosen in addition to the marginal models for the node and its parents. 

The proposed approach uses data reduction in two ways, first by the choice of the graph and second by the forward selection of parent nodes ordered by importance. Edges pointing to a node in the starting graph can be removed when in the D-vine regression of the node the corresponding parent node is not selected as a covariate. This allows for an edge selection without relying on the Gaussian assumption of the data. This approach identified three edges
in the network, which are not no longer supported by the experimental conditions investigated. The removal of these edges were found to be plausible given the specific experimental conditions.

We illustrated the approach using the $cd3cd28+aktinhib$ experimental setting of the Sachs data. Here we showed a much better fit of the copula based SEM with margins fitted by mixtures of univariate normals or nonparametric kernel density estimation and using non Gaussian pair copulas. 

These D-vine copula based SEMs are easy to fit and can be applied to large networks, since D-vine regressions are feasible for several hundred covariates. In addition it is easy to simulate from the model and assess conditional densities or probabilities of the nodes given observed values of the selected parent nodes.

These D-vine copula based SEM's can also be utilized to conduct a causal analysis using the pooled Sachs data. For this we would subdivide the data in a node specific way. For a given node we would select only observations from experimental settings which do not inhibit or activate the node. This would remove the interventional effects present in the data and such an analysis is currently investigated.

While it seems like that the knowledge of an initial graph is restrictive, the approach can be combined with any structure learning approach. In particular to allow for non Gaussian structure selection the vine copula based approach of \cite{bauer2012pair} to test for conditional independence in a PC algorithm
\citep{kalisch2007estimating, spirtes2011causation} can be used and the selected graph can be fitted using the proposed D-vine copula based SEM.

Two immediate extensions of the proposed modeling approach are to allow for R-vine based SEMs or to improve the forward selection procedure by looking two steps ahead instead of a step as has been proposed by \cite{tepegjozova2021nonparametric}.

A non Gaussian analysis of the complete Sachs data would require the development of efficient mixtures of vine regression models to accommodate the different clusters induced by the different experimental settings.  A first step in this direction is \cite{sahin2021vine}, who developed a data driven algorithm for the specification of vine distributions in a mixture setup for clustering. Such a development might further improve the fit of the $cd3cd28+aktinhib$ experimental, since for some nodes we can observe clusters in the data (see Figure \ref{fig:pairsPlotsMarg}).










%% file: 7-Appendix.tex
\clearpage

\section*{Appendix}
\beginappendix

\subsection*{A1: Marginal fits for the copula based structural equation model}
 \begin{figure}[ht]
    \centering
    \subfloat[\textbf{M}$^{\mathbf{ker}}$]{{\includegraphics[scale=0.12]{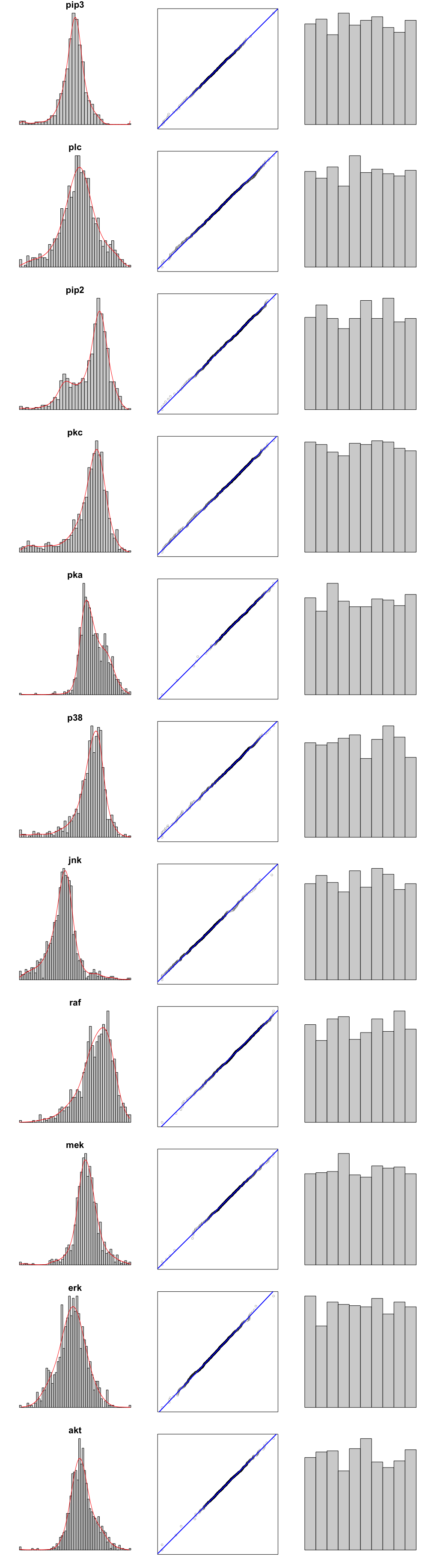}}}%
    \qquad
    \subfloat[\textbf{M}$^{\mathbf{mix}}$]{{\includegraphics[scale=0.12]{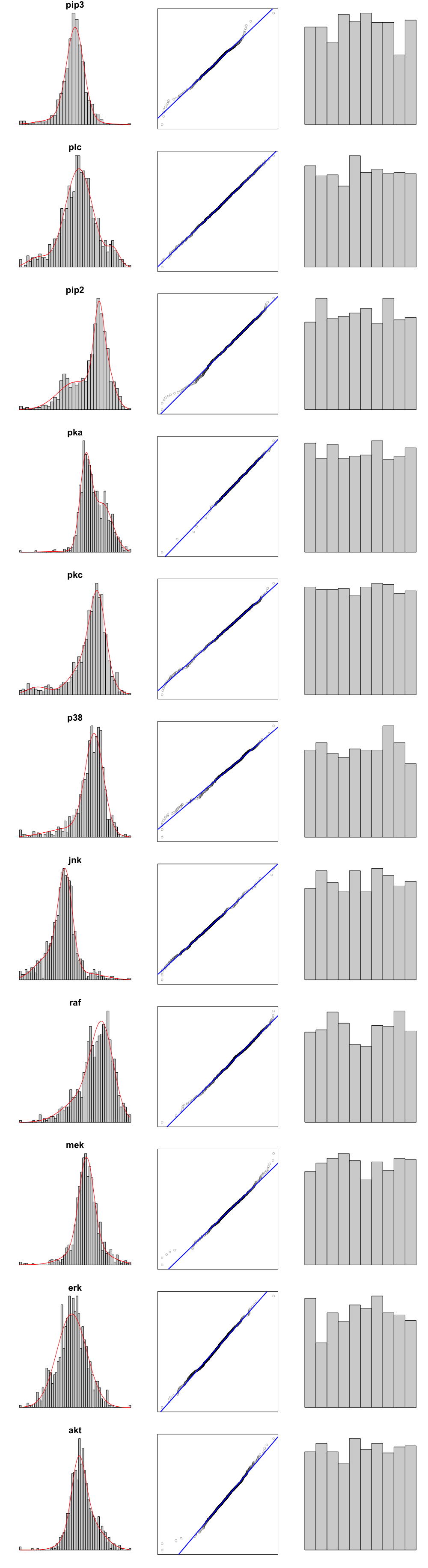}}}
    \qquad
    \subfloat[\textbf{M}$^{\mathbf{gauss}}$]{{\includegraphics[scale=0.12]{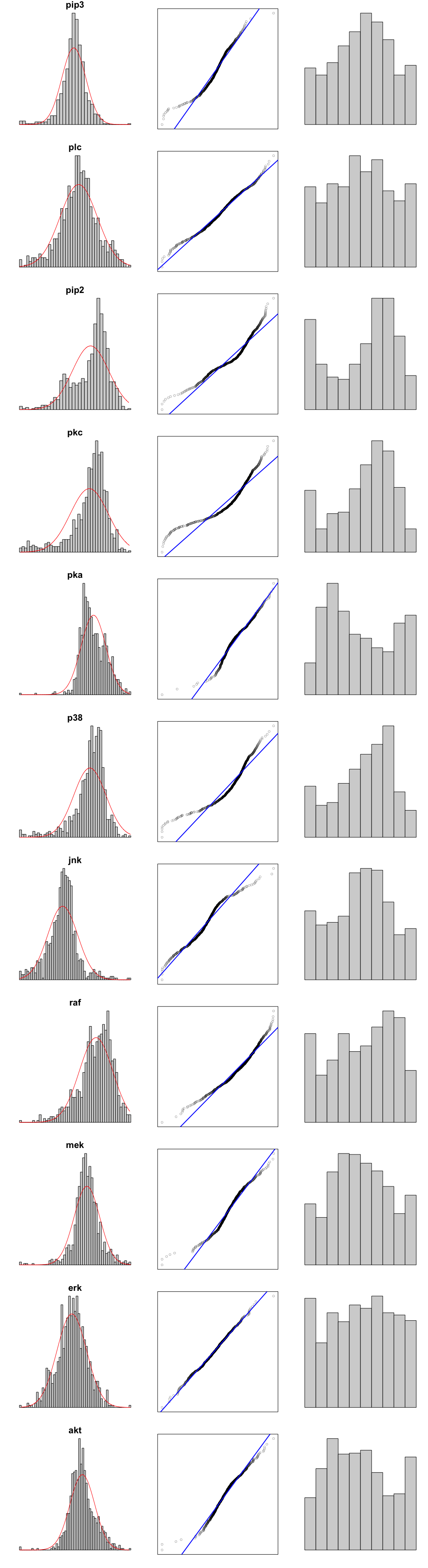}}}%
    \caption{Density plot and qq plot of the chosen margins (in the first two columns) and histogram of the data after applying the distribution function of the fitted marginal distributions, in the third column, for all nodes for the chosen marginals on the $cd3cd28+aktinhib$ data set}%
    \label{fig:contALL3X3}
\end{figure}

\begin{table}[ht]
\resizebox{\columnwidth}{!}{
\centering
\tiny
\begin{tabular}{|r||r|r|r|r||r|r|r|r||r|r|r|}
 \hline
&\multicolumn{4}{c||}{Kernel density estimates}&\multicolumn{4}{c||}{
Gaussian mixture margins}&\multicolumn{3}{c|}{Gaussian margins}\\
  \hline
 Node  & Log-lik. & AIC$_M$ & BIC$_M$ & \# Par. & Log-lik. & AIC$_M$ & BIC$_M$ &\# Par. & Log-lik. & AIC$_M$ & BIC$_M$\\ 
  \hline
pip3 &  -895.20 & 1813.47 & 1868.14 & 11.54  & -916.83 & 1843.66 & 1867.36&5 & -986.90 & 1977.81 & 1987.29\\ \hline 
plc &  -952.81 & 1927.03 & 1977.73 & 10.70  & -951.96 & 1919.92 & 1957.83&8 & -979.48 & 1962.97 & 1972.44\\ \hline
pip2 &  -1214.93 & 2447.95 & 2490.80 & 9.04  & -1225.04 & 2466.07 & 2503.99&8 & -1338.18 & 2680.35 & 2689.83\\ \hline  
pkc &  -771.76 & 1558.08 & 1592.58 & 7.28 &  -777.10 & 1570.20 & 1608.11&8 & -954.10 & 1912.21 & 1921.69\\ \hline
pka &  -767.31 & 1552.31 & 1594.24 & 8.85 &  -769.54 & 1555.08 & 1592.99&8 & -831.22 & 1666.44 & 1675.92\\ \hline  
p38 &  -765.47 & 1546.82 & 1584.45 & 7.94 &  -778.79 & 1567.58 & 1591.28&5 & -925.20 & 1854.40 & 1863.88\\ \hline
jnk &  -923.74 & 1874.36 & 1938.03 & 13.44 &  -929.73 & 1875.46 & 1913.37&8 & -1008.16 & 2020.31 & 2029.79\\ \hline
raf &  -921.52 & 1865.16 & 1917.58 & 11.06  & -926.74 & 1863.48 & 1887.18& 5&-971.98 & 1947.95 & 1957.43\\ \hline
mek &  -710.24 & 1450.37 & 1521.20 & 14.94  & -725.23 & 1460.47 & 1484.16&5& -785.21 & 1574.41 & 1583.89\\ \hline
erk &  -991.42 & 1997.43 & 2032.01 & 7.30 &  -997.54 & 1999.07 & 2008.55&2& -997.54 & 1999.07 & 2008.55\\ \hline
akt &  -865.88 & 1753.80 & 1806.03 & 11.02 &  -880.65 & 1771.30 & 1794.99&5 & -912.43 & 1828.85 & 1838.33\\ \hline
$\sum$: & -9780.28 & 19786.78 & 20322.79 & 113.11  & -9879.15 & 19892.29 & 20209.81 &67& -10690.40 & 21424.77 & 21529.04\\ 
   \hline
\end{tabular}
}
\caption{Marginal log-likelihood, AIC$_M$ and BIC$_M$ of the fitted margins. For the fitted kernel density estimates and for the fitted Gaussian mixtures margins the number of (effective) parameters is displayed}
\label{tab:Margins}
\end{table}
We see that all three measures are extremely similar for the fitted Gaussian mixture margins and kernel density estimates. For the log-likelihood and the AIC the kernel density estimates are slightly better whereas for the BIC the Gaussian mixture margins marginally outperform kernel density estimates. This difference is due to the higher number of parameters necessary to fit kernel density estimates. Compared to these two approaches fitting Gaussian margins results overall in a worse fit. 
\clearpage 
\subsection*{A2: Exploring pairwise dependence under different marginal specifications}
\label{sec:pairexplore}
 \begin{figure}[ht!]
    \centering
    \subfloat[Kernel density estimates]{{\includegraphics[scale=0.25]{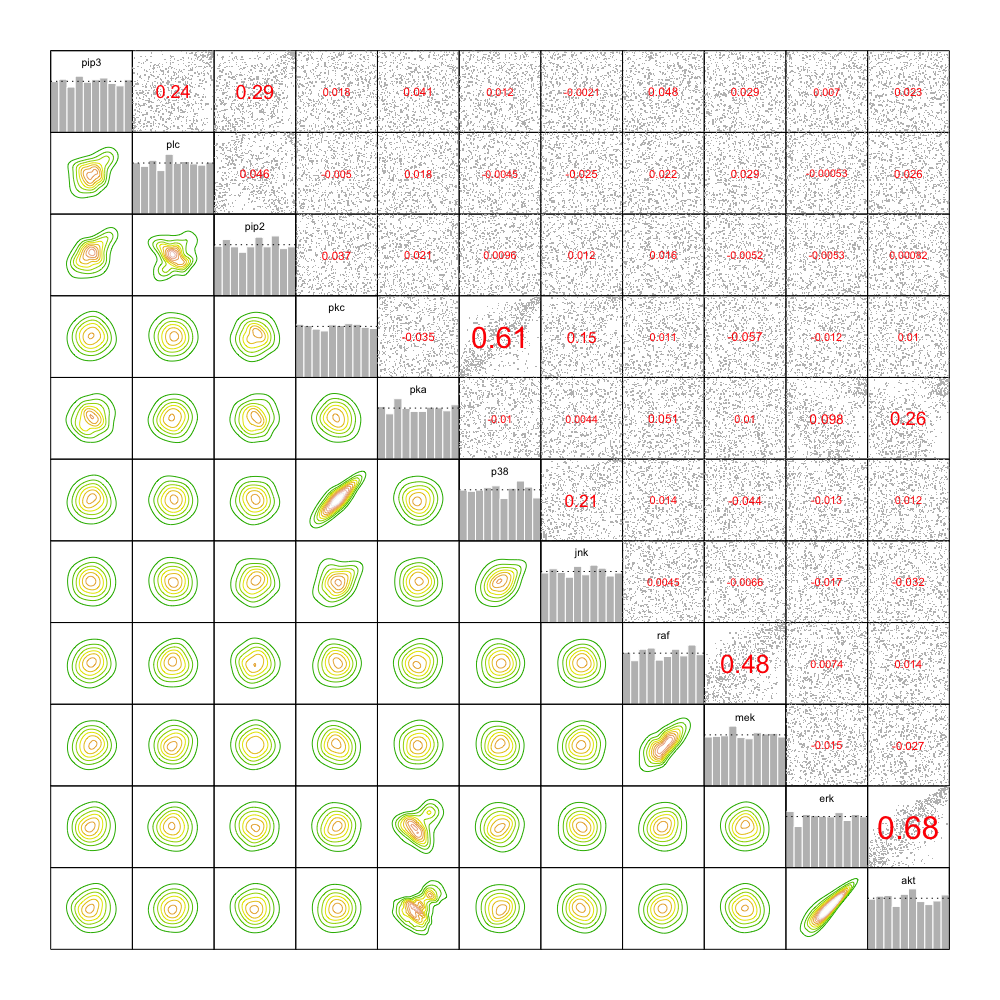}}}\\
    \subfloat[Gaussian mixture margins]{{\includegraphics[scale=0.25]{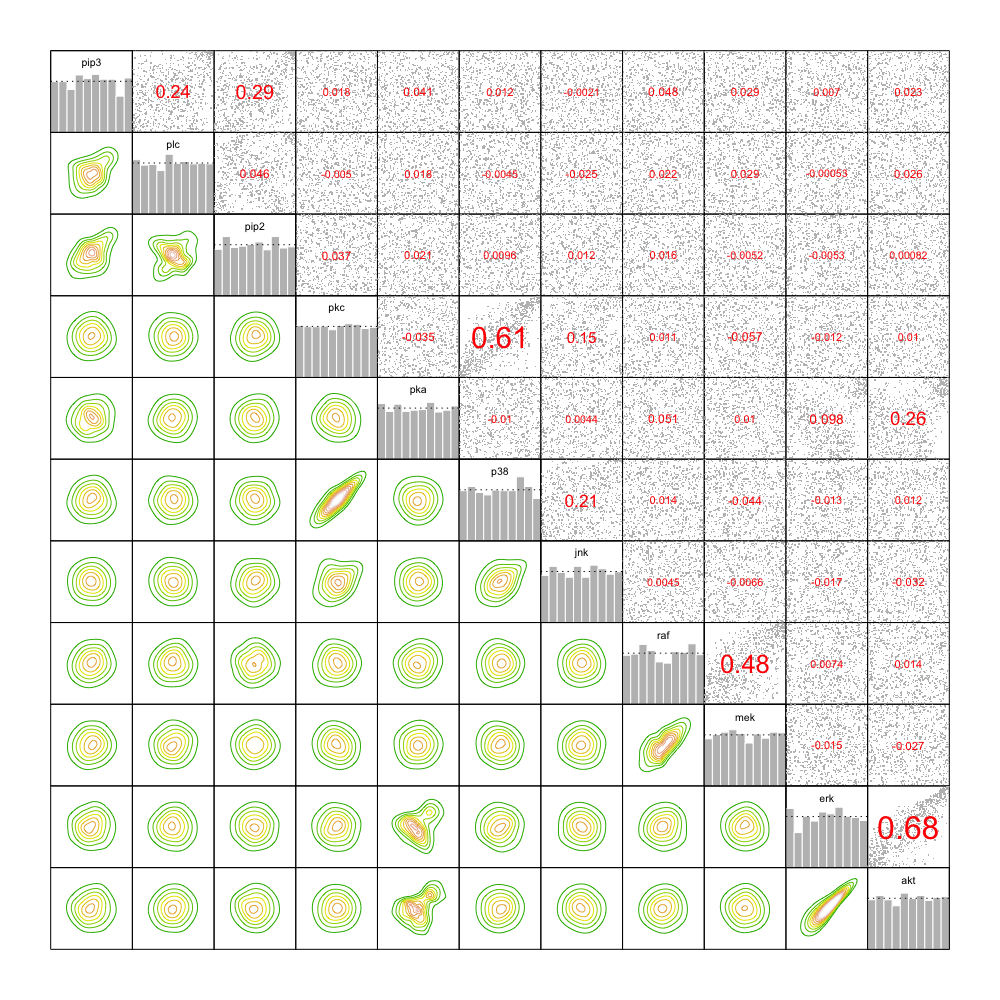}}}\\
\end{figure}
\newpage
 \begin{figure}
 \centering
 \setcounter{subfigure}{2}
    \subfloat[Gaussian margins]{{\includegraphics[scale=0.25]{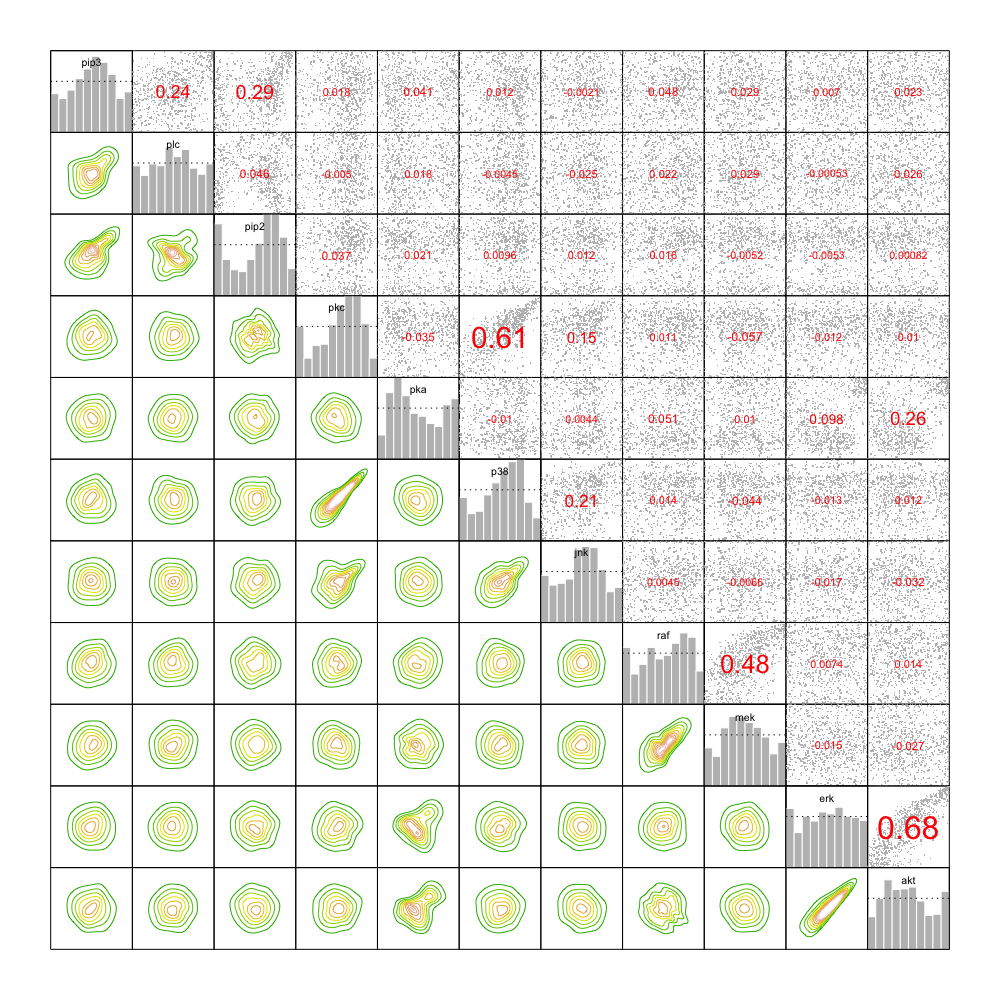}}}\\
    \caption{Normalized contour plots in the lower left triangle, normalized scatter plots in the upper right triangle and histograms in the diagonal elements after applying the PIT using the different fitted margins to the  $cd3cd28+aktinhib$ experiment}%
            \label{fig:pairsPlotsMarg}
\end{figure}
\clearpage

\subsection*{A3: Copula fits for the copula based SEM}
\label{sec:copulafit}
\vspace*{-.5cm}

\begin{table}[ht]
\resizebox{\columnwidth}{!}{
\centering
\tiny
\begin{tabular}{|r||r|r|r|r||r|r|r|r||r|r|r|r|}
 \hline
&\multicolumn{4}{c||}{\textbf{M}$^{\mathbf{ker}}$\textbf{C}$^{\mathbf{pnp}}$}&\multicolumn{4}{c||}{
\textbf{M}$^{\mathbf{mix}}$\textbf{C}$^{\mathbf{pnp}}$}&\multicolumn{4}{c|}{\textbf{M}$^{\mathbf{gauss}}$\textbf{C}$^{\mathbf{gauss}}$}\\
  \hline
 Node  & Order& Log-lik. & AIC$_C$ & BIC$_C$ &Order& Log-lik. & AIC$_C$ & BIC$_C$ & Order&Log-lik. & AIC$_C$ & BIC$_C$\\ 
  \hline
plc & pip3 & 126.09 & -209.26 & -107.55 & pip3 & 125.75 & -209.73 & -110.76 &pip3 & 57.72 & -113.43 & -108.69\\  \hline
pip2 & pip3, plc & 293.69 & -509.66 & -325.46  & pip3, plc & 290.73 & -498.59 & -302.22&pip3 & 88.80 & -175.61 & -170.87 \\  \hline  
pkc & pip2 & 1.82 & -1.64 & 3.09 & pip2 & 1.84 & -1.68 & 3.06 & pip2 & 1.73 & -1.46 & 3.28\\  \hline  
pka & pkc & 1.08 & -0.16 & 4.58 & pkc & 1.13 & -0.26 & 4.48 &  & 0 & 0 & 0 \\  \hline p38 & pkc, pka & 519.59 & -982.02 & -846.57 & pkc, pka & 515.32 & -977.28 & -850.79 & pkc & 354.14 & -706.28 & -701.54\\  \hline
jnk & pkc, pka & 86.66 & -124.26 & -7.99 & pkc, pka & 84.79 & -128.34 & -30.61 & pkc & 15.54 & -29.08 & -24.34\\  \hline 
raf & pka, pkc & 20.00 & -6.88 & 71.59 & pka, pkc & 19.32 & -9.59 & 59.28 &pka, pkc & 2.15 & -0.30 & 9.18\\  \hline
mek & raf, pkc, pka & 310.82 & -574.83 & -463.88 & raf, pkc, pka & 308.73 & -571.78 & -463.55 &raf, pkc & 232.86 & -461.72 & -452.24 \\  \hline
erk & pka & 169.47 & -302.22 & -215.22 & pka & 167.42 & -297.22 & -208.06 & pka & 29.16 & -56.33 & -51.59\\  \hline
akt & erk, pka & 787.00 & -1517.80 & -1384.65& erk, pka & 777.31 & -1495.02 & -1353.81 & erk, pka & 715.74 & -1427.47 & -1417.99\\  \hline
$\sum$:&  & 2316.23 & -4228.73 & -3272.06 &  & 2292.34 & -4189.49 & -3252.98 &  & 1497.84 & -2971.68 & -2914.80\\ 
   \hline
\end{tabular}
}
\caption{Log-likelihood of the copula terms, AIC$_C$, BIC$_C$ and order of the D-vine for each node in the D-vine regression models}
\label{tab:Cops}
\end{table}

\vspace*{-1cm}
\begin{table}[ht]
  \centering
  \tiny
\begin{tabular}{|r|r|r|r|r|r|r|r|r|}
  \hline
  &&&Effective&Copula&&&\\
(a) \textbf{M}$^{\mathbf{ker}}$\textbf{C}$^{\mathbf{pnp}}$& Pair copula & Family & \# parameters & log-lik. & AIC$_C$ & BIC$_C$ & Est. Ken. $\tau$ \\ 
  \hline  \hline
  plc &  plc pip3 & tll & 21.46 & 126.09 & -209.26 & -107.55 & 0.24 \\       \hline
  \multirow{2}{*}{pip2} & pip2 pip3 & tll & 19.77 & 175.60 & -311.66 & -217.98 & 0.05 \\ 
   &  pip2 plc; pip3 & tll & 19.10 & 118.10 & -198.00 & -107.48 & 0.29 \\   \hline
  pkc & pkc pip2 & clayton & 1.00 & 1.82 & -1.64 & 3.09 & -0.01 \\   \hline
  pka & pka pkc & frank & 1.00 & 1.08 & -0.16 & 4.58 & -0.03 \\    \hline
  \multirow{2}{*}{p38} & p38 pkc & tll & 27.58 & 517.89 & -980.62 & -849.91 & 0.60 \\  
   & p38 pka; pkc & joe & 1.00 & 1.70 & -1.40 & 3.34 & -0.01 \\      \hline
  \multirow{2}{*}{jnk} & jnk pkc & tll & 23.53 & 85.59 & -124.11 & -12.58 & 0.15 \\ 
   &  jnk pka; pkc & gumbel & 1.00 & 1.08 & -0.15 & 4.59 & 0.00 \\ 
   \hline
\multirow{2}{*}{raf} & raf pka & tll & 15.56 & 18.74 & -6.36 & 67.37 & 0.05\\
   & raf pkc; pka & clayton & 1.00 & 1.26 & -0.52 & 4.22 & 0.01 \\       \hline
  \multirow{3}{*}{mek} & mek raf & tll & 21.41 & 304.00 & -565.19 & -463.72 & 0.48 \\
   & mek pkc; raf & gaussian & 1.00 & 4.54 & -7.09 & -2.35 & -0.06 \\ 
   &  mek pka; raf pkc & gumbel & 1.00 & 2.28 & -2.55 & 2.19 & 0.01 \\      \hline
  erk& erk pka & tll & 18.36 & 169.47 & -302.22 & -215.22 & 0.10 \\     \hline
  \multirow{2}{*}{akt} & akt erk & tll & 26.10 & 664.89 & -1277.59 & -1153.92 & 0.67 \\
   & akt pka; erk & bb8 & 2.00 & 122.11 & -240.21 & -230.73 & 0.26 \\   \hline
\end{tabular}\\
\vspace*{.25cm}

\begin{tabular}{|r|r|r|r|r|r|r|r|r|}
  \hline
  &&&Effective&Copula&&&\\
(b) \textbf{M}$^{\mathbf{mix}}$\textbf{C}$^{\mathbf{pnp}}$ & Pair copula & Family & \# parameters & log-lik. & AIC$_C$ & BIC$_C$ & Est. Ken. $\tau$ \\ 
  \hline  \hline
  plc &  plc pip3 & tll & 20.88 & 125.75 & -209.73 & -110.76 & 0.24 \\        \hline
  \multirow{2}{*}{pip2} & pip2 pip3 & tll & 20.32 & 174.14 & -307.64 & -211.35 & 0.05 \\ 
   &  pip2 plc; pip3 & tll & 21.12 & 116.59 & -190.95 & -90.87 & 0.29 \\    \hline
  pkc & pkc pip2 & clayton & 1.00 & 1.84 & -1.68 & 3.06 & -0.01 \\   \hline
  pka & pka pkc & frank & 1.00 & 1.13 & -0.26 & 4.48 & -0.03 \\    \hline
  \multirow{2}{*}{p38} & p38 pkc & tll & 25.69 & 513.63 & -975.89 & -854.14 & 0.60 \\  
   & p38 pka; pkc & joe & 1.00 & 1.69 & -1.39 & 3.35 & -0.01 \\       \hline
  \multirow{2}{*}{jnk} & jnk pkc & tll & 19.62 & 83.69 & -128.13 & -35.14 & 0.15 \\ 
   &  jnk pka; pkc & gumbel & 1.00 & 1.10 & -0.21 & 4.53 & 0.00 \\ 
   \hline
\multirow{2}{*}{raf} & raf pka & tll & 13.53 & 18.11 & -9.16 & 54.97 & 0.05 \\ 
   & raf pkc; pka & clayton & 1.00 & 1.21 & -0.43 & 4.31 & 0.01 \\        \hline
  \multirow{3}{*}{mek} & mek raf & tll & 20.84 & 301.89 & -562.11 & -463.35 & 0.48 \\
   & mek pkc; raf & gaussian & 1.00 & 4.67 & -7.33 & -2.59 & -0.06 \\ 
   &  mek pka; raf pkc & gaussian & 1.00 & 2.17 & -2.34 & 2.39 & 0.01 \\      \hline
  
  erk& erk pka & tll & 18.81 & 167.42 & -297.22 & -208.06 & 0.10 \\      \hline
  \multirow{2}{*}{akt} & akt erk & tll & 27.80 & 652.93 & -1250.27 & -1118.54 & 0.67 \\ 
   & akt pka; erk & bb8 & 2.00 & 124.38 & -244.75 & -235.27 & 0.26 \\    \hline
\end{tabular}\\
\vspace*{.25cm}

\begin{tabular}{|r|r|r|r|r|r|r|r|r|r|}
  \hline
 &&& Copula&&&&\\
(c) \textbf{M}$^{\mathbf{gauss}}$\textbf{C}$^{\mathbf{gauss}}$ & Pair copula & Family  & log-lik. & AIC$_C$ & BIC$_C$ & Par. & Est. Ken. $\tau$ \\ 
  \hline  \hline
  plc&plc pip3 & gaussian & 57.72 & -113.43 & -108.69 & 0.36 & 0.24 \\ \hline
   pip2 &pip2 pip3 & gaussian & 88.80 & -175.61 & -170.87 & 0.44 & 0.05 \\ \hline
  pkc & pkc pip2 & gaussian & 1.73 & -1.46 & 3.28 & 0.06 & -0.01 \\    \hline
  p38 & p38 pkc & gaussian & 354.14 & -706.28 & -701.54 & 0.75 & 0.60 \\  \hline
jnk & jnk pkc & gaussian & 15.54 & -29.08 & -24.34 & 0.19 & 0.15 \\ 
   \hline
 \multirow{2}{*}{raf} & raf pka & gaussian & 1.08 & -0.15 & 4.59 & 0.05 & 0.05 \\ 
   & raf pkc; pka & gaussian & 1.08 & -0.15 & 4.59 & 0.05 & 0.01 \\   \hline
   \multirow{2}{*}{mek} & mek raf & gaussian & 227.58 & -453.17 & -448.43 & 0.65 & 0.48 \\
   & mek pkc; raf & gaussian & 5.27 & -8.55 & -3.81 & -0.11 & -0.06 \\   \hline  
   erk & erk pka & gaussian & 29.16 & -56.33 & -51.59 & 0.26 & 0.10 \\     \hline
   \multirow{2}{*}{akt} & akt erk & gaussian & 530.82 & -1059.63 & -1054.89 & 0.85 & 0.67 \\ 
   &  akt pka; erk & gaussian & 184.92 & -367.84 & -363.10 & 0.60 & 0.26 \\  \hline
\end{tabular} \\
\vspace*{-.25cm}
 \caption{Summary of all copulas fitted in the different D-vine regression models}
 \label{tab:CallMkerCopX3}
\end{table}

  \begin{figure}[ht]%
    \centering
    \subfloat[\textbf{M}$^{\mathbf{ker}}$\textbf{C}$^{\mathbf{pnp}}$]{{\includegraphics[scale=0.07]{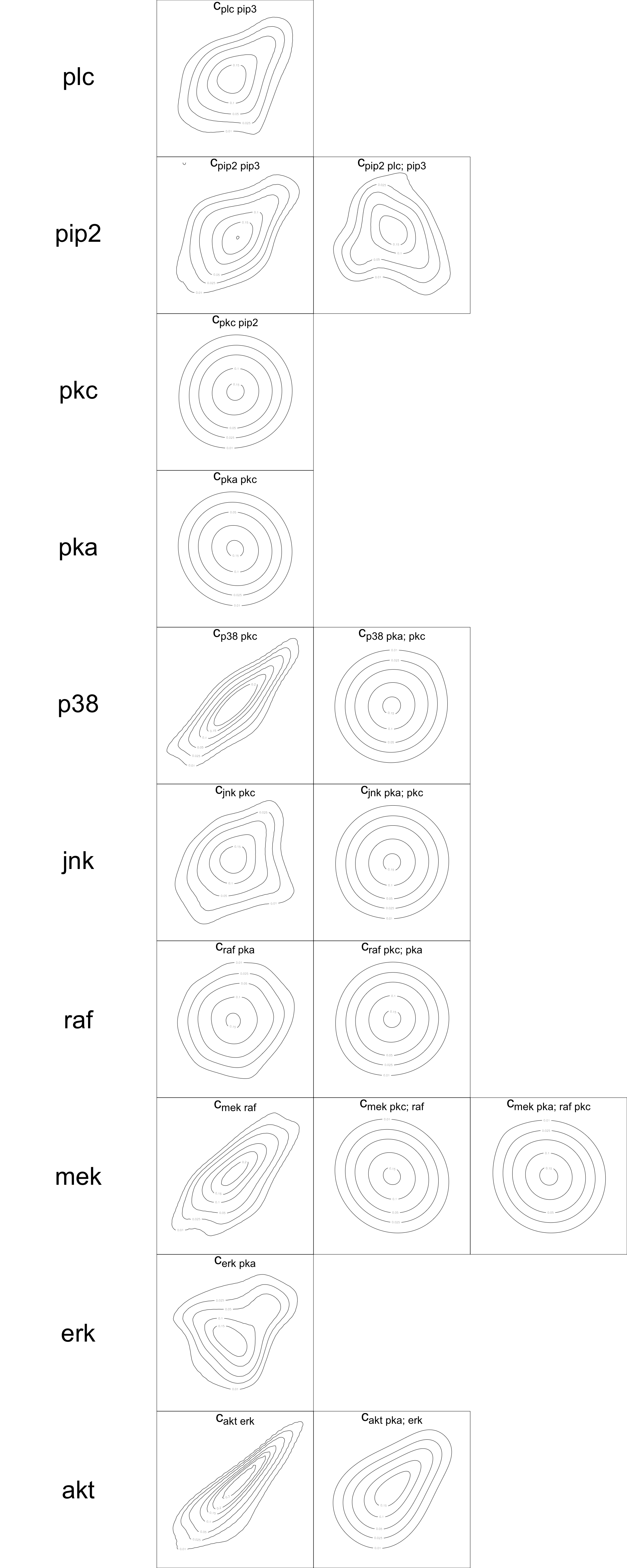}}}%
    \qquad
    \subfloat[\textbf{M}$^{\mathbf{mix}}$\textbf{C}$^{\mathbf{pnp}}$]{{\includegraphics[scale=0.07]{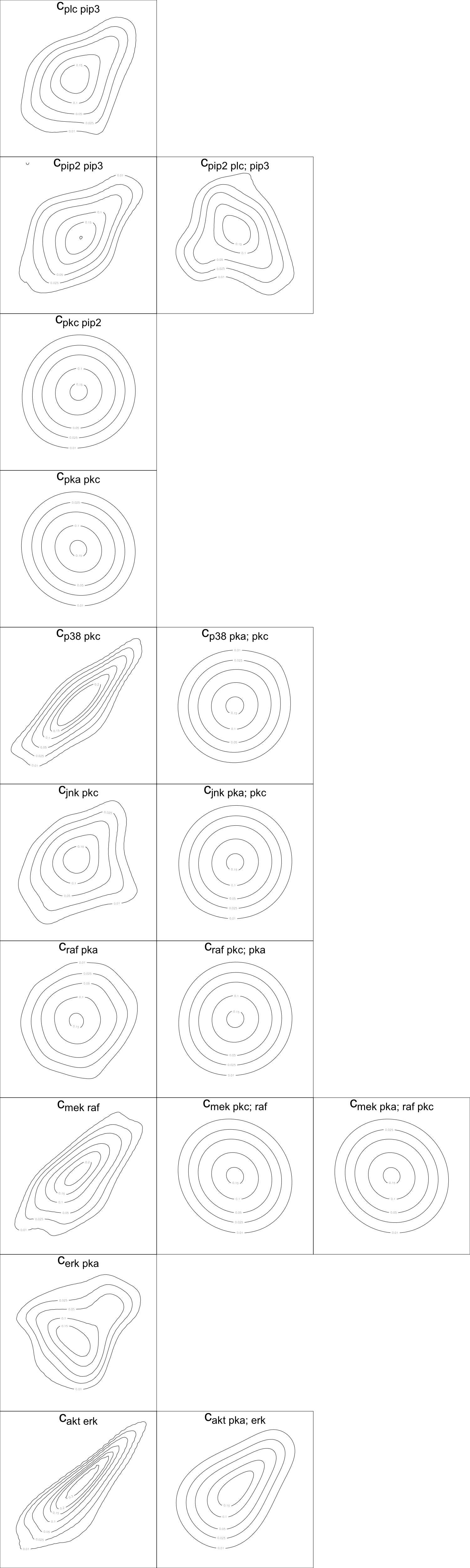}}}
    \qquad
    \subfloat[\textbf{M}$^{\mathbf{gauss}}$\textbf{C}$^{\mathbf{gauss}}$]{{\includegraphics[scale=0.07]{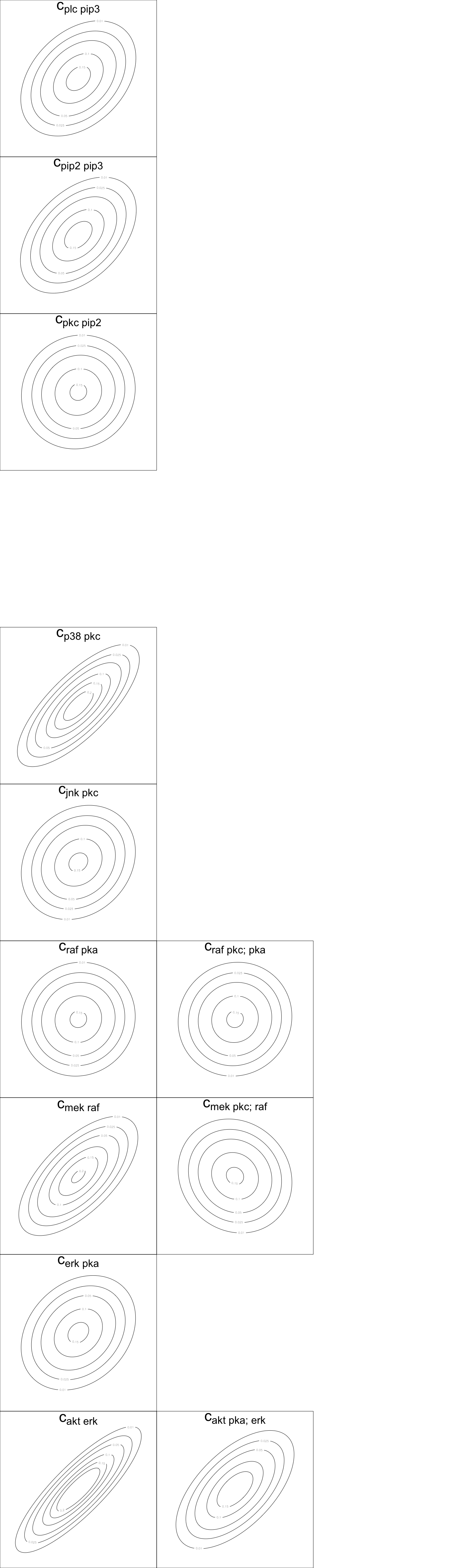}}}%
    \caption{Normalized contour plots for $cd3cd28+aktinhib$ data set under three marginal specifications}%
    \label{fig:contX3c}
\end{figure}
\clearpage

\bibliography{main}
\newpage

\section*{Supplementary Material}
\beginsupplement
\subsection*{S1: Comparison of linear Gaussian Bayesian network and structural equation model based on a Gaussian copula with Gaussian margins 
}
\label{sec:LGBN.Gauss}
\begin{table}[ht!]
\parbox{.45\linewidth}{
\centering
\tiny
\begin{tabular}{|c|c|c|c|}
  \hline
 Node & $\hat\beta_0$& $\hat{\bm\beta}$& $\hat\sigma$ \\ 
  \hline  \hline
  pip3&  3.52&                  &0.61\\ \hline
     plc&  1.20  &             0.35 & 0.52\\ \hline
  pip2&  2.28&         (0.00, 0.66)  &1.13\\ \hline
  pkc & 2.59&        (0.04, -0.01)  &0.56\\ \hline
  pka&   6.42&              -0.02 & 0.42\\ \hline
  p38 & 1.02 &       (0.02, 0.73) & 0.23\\ \hline
  jnk & 1.41&        (0.05, 0.20) & 0.61\\ \hline
  raf & 3.28&       (0.06, 0.05)  &0.58\\ \hline
   mek & 1.62&  (0.52, -0.03, -0.07) & 0.22\\ \hline
   erk & 0.98 &      (-0.03, 0.31)  &   0.58\\ \hline
   akt & -0.71 &(0.69, 0.36, 0.02)  &0.09\\ \hline
\end{tabular}
\caption*{(a) LGBN}
}
\hfill
\parbox{.45\linewidth}{
\centering
\tiny
\begin{tabular}{|c|c|c|c|}
  \hline
 Node & $\hat\beta_0$& $\hat{\bm\beta}$& $\hat\sigma$ \\ 
  \hline  \hline
  pip3&  3.52&                  &0.61\\ \hline
  plc&  1.20  &             0.35 & 0.52\\ \hline
  pip2&  2.28&         0.66  &1.13\\ \hline
  pkc & 2.57&        0.04  &0.56\\ \hline
  pka&   6.35&               & 0.42\\ \hline
  p38 & 1.15 &       0.73 & 0.23\\ \hline
  jnk & 1.73&        0.20 & 0.61\\ \hline
	raf & 3.29&       (0.06, 0.05)  &0.58\\ \hline
   mek & 1.42&  (0.52, -0.07) & 0.22\\ \hline
  erk & 0.88 &      0.31  &   0.58\\ \hline
   akt & -0.66 &(0.69, 0.36)  &0.09\\ \hline
\end{tabular}
\caption*{(b) \textbf{M}$^{\mathbf{gauss}}$\textbf{C}$^{\mathbf{gauss}}$}}

\caption{Estimated parameters of the conditional normal distributions $N(\beta_0+\bm{\beta}^\top x;\sigma^2)$ of each node on the set of its parents modeled in the linear Gaussian Bayesian network and the \textbf{M}$^{\mathbf{gauss}}$\textbf{C}$^{\mathbf{gauss}}$}
\label{tab:LGBNCOND}
\end{table}

\subsection*{S2: Chosen values of parent nodes}
\label{sec:conditioning}
\begin{figure}[h!]
\centering
\includegraphics[width=0.66\textwidth]{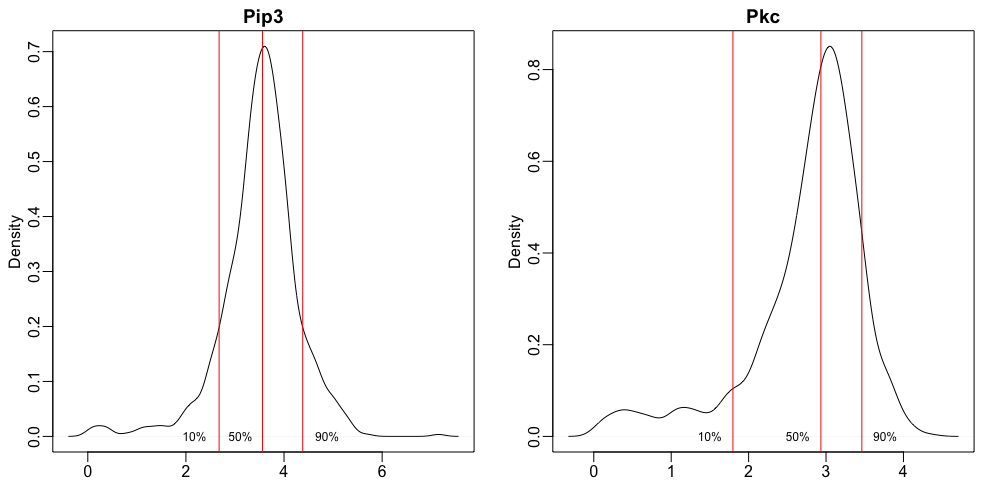}
\caption{Empirical density of the nodes $pip3$ and $pkc$ with horizontal lines at the $10\%$, $50\%$ and $90\%$ quantile}
\label{fig:1dim}
\end{figure}

\begin{figure}[ht!]
\centering
\includegraphics[width=0.66\textwidth]{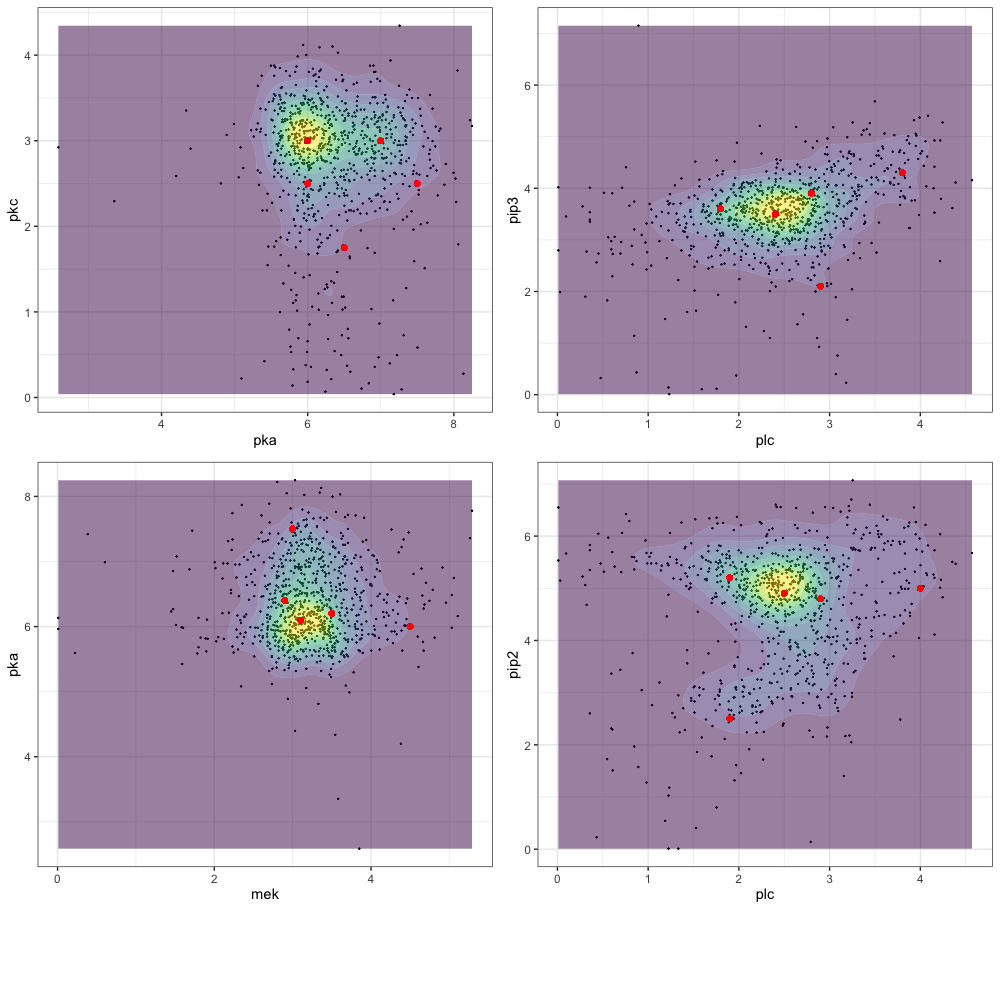}
\caption{Contour plots of the two-dimensional kernel density estimates of the node pairs $pka\leftrightarrow pkc$, $plc\leftrightarrow pip3$, $mek\leftrightarrow pka$ and $plc\leftrightarrow pip2$. Red points correspond to the points chosen for conditioning values of the parents}
\label{fig:2dim}
\end{figure}

  \begin{figure}[!ht]
    \centering
    \subfloat{{\includegraphics[scale=0.17]{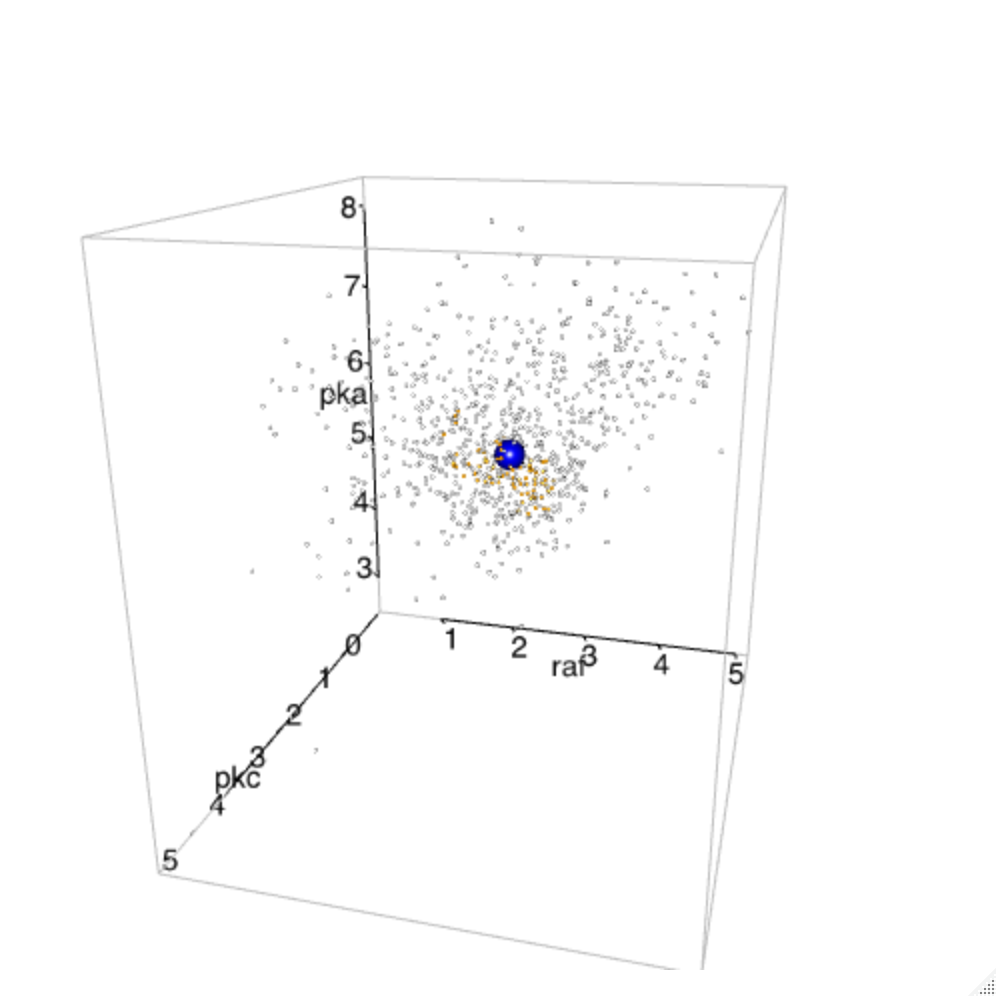}}}%
    \subfloat{{\includegraphics[scale=0.17]{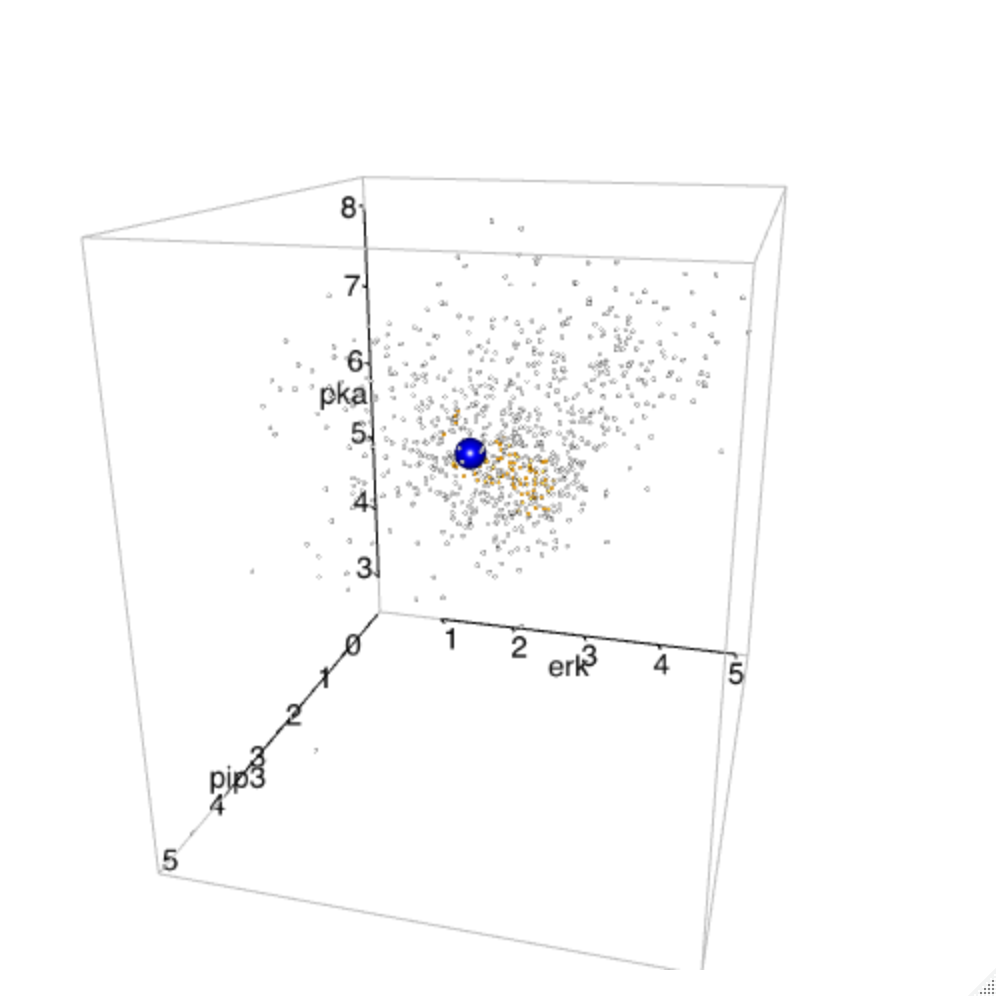}}}
    \bigskip
    
    \subfloat{{\includegraphics[scale=0.17]{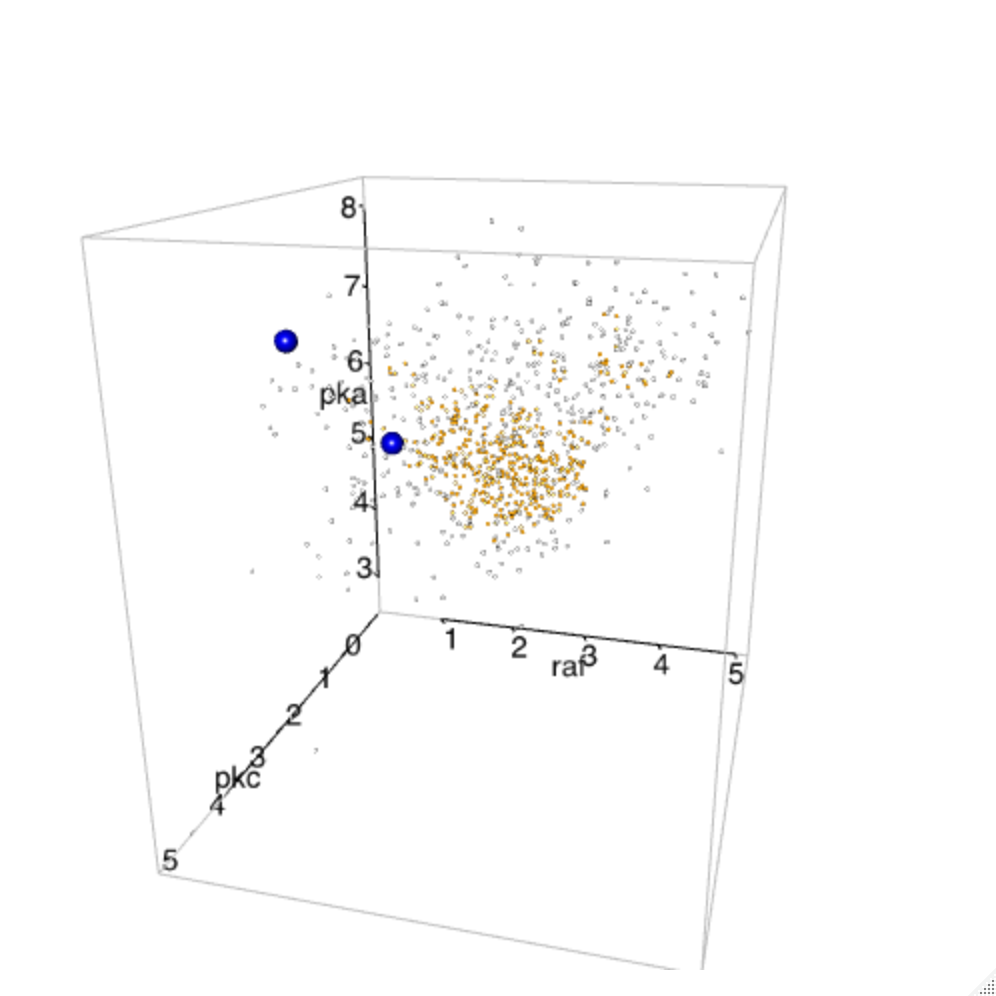}}}%
    \subfloat{{\includegraphics[scale=0.17]{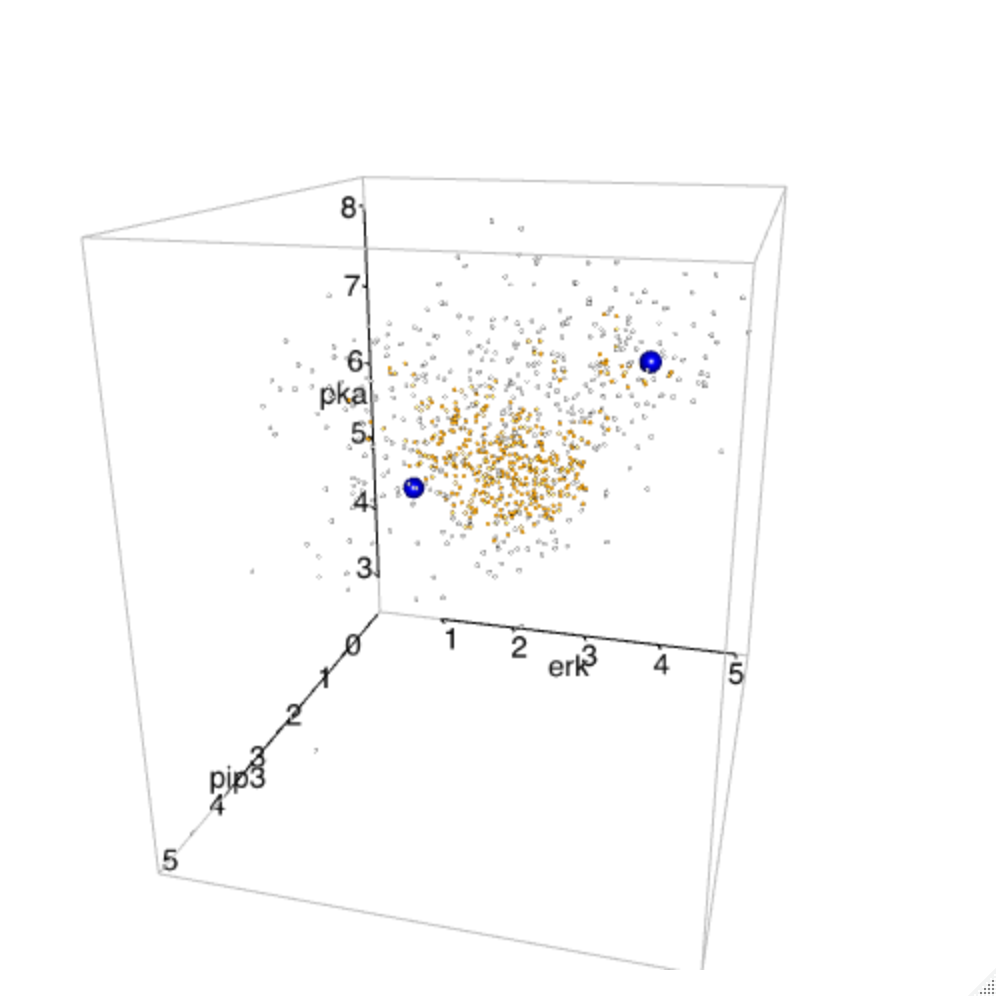}}}
    \bigskip
    
        \subfloat{{\includegraphics[scale=0.17]{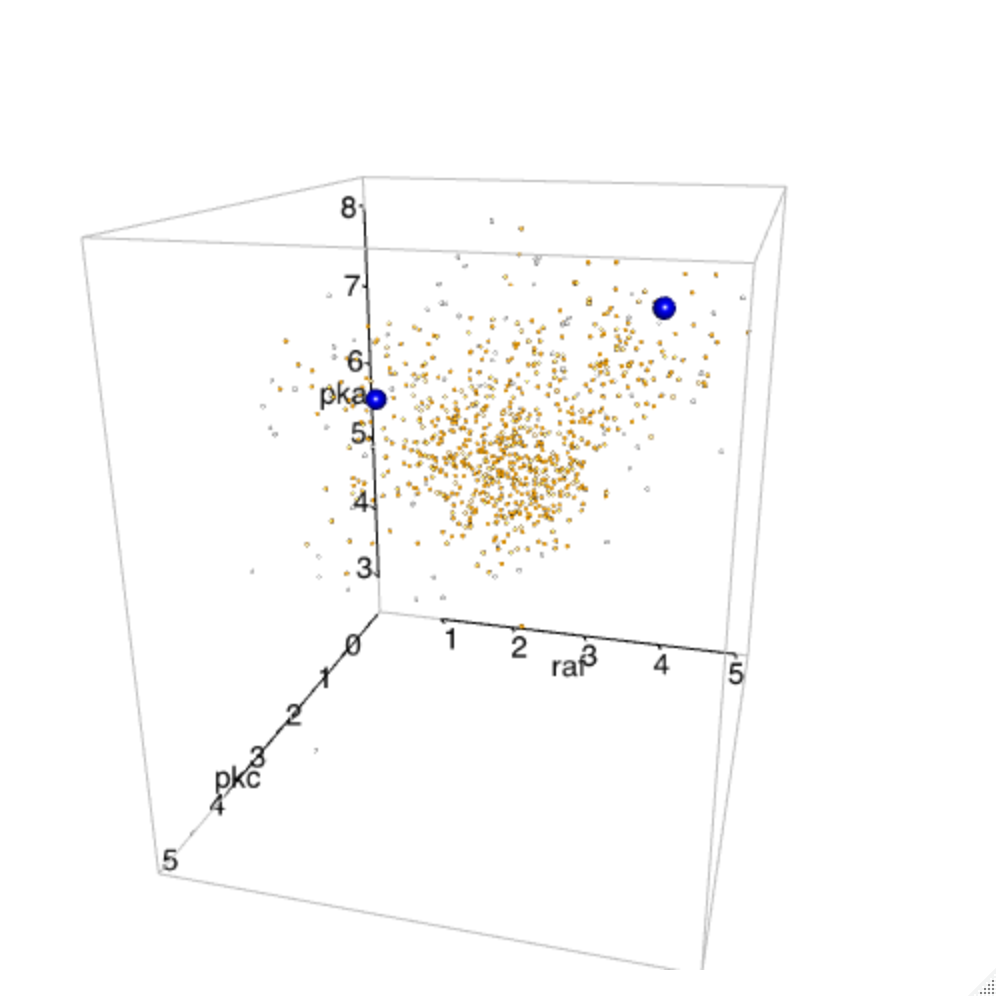}}}%
    \subfloat{{\includegraphics[scale=0.17]{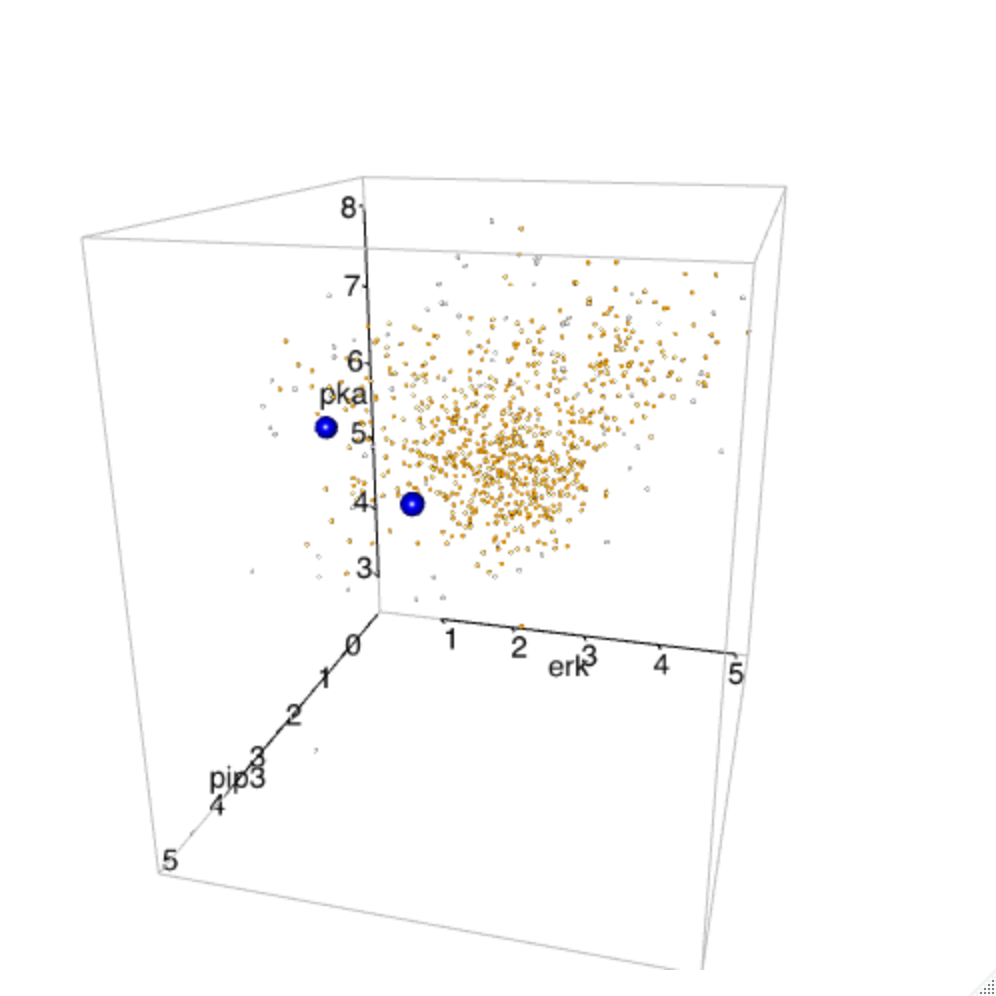}}}
    \bigskip
    
    \caption{Scatter plots of the parents of the nodes $mek$ (left) and $akt$ (right). Points in yellow are points within the $10\%$, $50\%$ and $90\%$ quantile of the fitted three-dimensional kernel density estimates (from top to bottom). Blue points are the points chosen for conditioning}%
    \label{fig:3dim}
\end{figure}
\clearpage